\documentclass[reqno,11pt]{amsart}

\usepackage{amsmath}
\usepackage{amssymb}

\newtheorem{Theorem}{Theorem}[section]
\newtheorem{Lemma}[Theorem]{Lemma}
\newtheorem{Proposition}[Theorem]{Proposition}
\newtheorem{Corollary}[Theorem]{Corollary}

\theoremstyle{definition}
\newtheorem{Remark}[Theorem]{Remark}
\newtheorem{Example}[Theorem]{Example}

\numberwithin{equation}{section}

\begin{document}

\title[Heat expansions on the integers and the Toda lattice]
{Heat kernel expansions on the integers and the Toda lattice hierarchy}

\author[P.~Iliev]{Plamen~Iliev}
\address{School of Mathematics, Georgia Institute of Technology, 
Atlanta, GA 30332--0160, USA}
\email{iliev@math.gatech.edu}

\date{December 6, 2007}

\begin{abstract}
We consider the heat equation $u_t=Lu$ where $L$ is a 
second-order difference operator in a discrete variable $n$. The fundamental 
solution has an expansion in terms of the Bessel functions of imaginary 
argument. The coefficients $\alpha_k(n,m)$ in this expansion are 
analogs of Hadamard's coefficients for the (continuous) Schr\"odinger 
operator. 

We derive an explicit formula for $\alpha_k$ in terms of the wave and 
the adjoint wave functions of the Toda lattice hierarchy. As a first 
application of this result, we prove that the values of these coefficients on 
the diagonals $n=m$ and $n=m+1$ define a hierarchy of differential-difference 
equations which is equivalent to the Toda lattice hierarchy. Using this fact 
and the correspondence between commutative rings of difference operators and 
algebraic curves we show that the fundamental solution can be summed up, 
giving a finite formula involving only two Bessel functions with polynomial 
coefficients in the time variable $t$, if and only if the operator $L$ belongs 
to the family of bispectral operators constructed in \cite{HI2}.
\end{abstract}

\maketitle

\newcommand{\thref}[1]{Theorem \ref{#1}}
\newcommand{\leref}[1]{Lemma \ref{#1}}
\newcommand{\prref}[1]{Proposition \ref{#1}}
\newcommand{\reref}[1]{Remark \ref{#1}}
\newcommand{\deref}[1]{Definition \ref{#1}}
\newcommand{\exref}[1]{Example \ref{#1}}
\newcommand{\coref}[1]{Corollary \ref{#1}}
\newcommand{\seref}[1]{Section \ref{#1}}
\newcommand{\ssref}[1]{Subsection \ref{#1}}

\newcommand{\cC}{{\mathcal C}}
\newcommand{\cA}{{\mathcal A}}
\newcommand{\cP}{{\mathcal P}}
\newcommand{\cQ}{{\mathcal Q}}
\newcommand{\cR}{{\mathcal R}}
\newcommand{\cS}{{\mathcal S}}
\newcommand{\cV}{{\mathcal V}}
\newcommand{\cW}{{\mathcal W}}

\newcommand{\C}{\mathbb C}
\newcommand{\Z}{\mathbb Z}
\newcommand{\N}{\mathbb N}
\newcommand{\X}{\mathbb X}
\newcommand{\Y}{\mathbb Y}

\newcommand{\fl}{\mathfrak{l}}
\newcommand{\fr}{\mathfrak{r}}
\newcommand{\ff}{\mathfrak{f}}
\newcommand{\fg}{\mathfrak{g}}
\newcommand{\fhh}{\mathfrak{h}}
\newcommand{\fp}{\mathfrak{p}}
\newcommand{\fq}{\mathfrak{q}}

\newcommand{\ft}{{\tilde{f}}}
\newcommand{\fh}{{\hat{f}}}

\newcommand{\de}{{\delta}}
\newcommand{\dep}{{\delta'}}
\newcommand{\ga}{{\gamma}}
\newcommand{\gap}{{\gamma'}}

\newcommand{\res}{\mathrm{res}}
\newcommand{\pd}{\partial}
\newcommand{\Id}{\mathrm{Id}}
\newcommand{\RHS}{\mathrm{RHS}}
\newcommand{\LHS}{\mathrm{LHS}}
\newcommand{\Spec}{\mathrm{Spec}}
\newcommand{\Wr}{\mathrm{Wr}_{\Delta}}

\newcommand{\Psib}{\bar{\Psi}}
\newcommand{\Psibs}{\bar{\Psi}^*}

\newcommand{\Qip}{Q_{\infty}^+}
\newcommand{\Qim}{Q_{\infty}^-}

\tableofcontents

\section{Introduction}\label{se1}

The fundamental solution of the heat equation
\begin{equation*}
u_t = u_{xx} + V(x)u,\quad u(x,0) = \delta_y(x)
\end{equation*}
has an asymptotic expansion of the form
\begin{equation*}
u(x,y,t) \sim \frac {e^{-\frac{(x-y)^2}{4t}}}{\sqrt{4\pi t}} \left( 1 +
\sum_{k=1}^{\infty} H_k(x,y)t^n\right) \text{ as }t\rightarrow 0+,
\end{equation*}
where the Hadamard's coefficients $H_n(x,y)$ are defined in a neighborhood of 
the diagonal $x=y$. Restricted on the diagonal, the coefficients $H_n(x,x)$ 
become differential polynomials of the potential $V(x)$ and can be used to 
define the Korteweg-de Vries (in short KdV) hierarchy. There are several 
proofs and applications of this fact, see for instance 
\cite{ASch,McKvM,Pol,Sch} 
and the references therein. One of the important steps in these papers is to 
use the connection between the heat kernel and the resolvent of the 
corresponding Schr\"odinger operator and the work of Gelfand and 
Dickey \cite{GD}. In particular, this construction was used to obtain explicit 
formulas for the higher KdV equations exploring different properties of 
Hadamard's coefficients. 

On the other hand, after the works \cite{DJKM,SS,SW}, we have a much 
better understanding of the KdV hierarchy and the parametrization of its 
solutions. Therefore it seems natural to reexamine the connection 
between the heat kernel and the KdV hierarchy within the framework of Sato's 
theory and to use the soliton technology 
as a tool to investigate the heat expansions. This idea was developed 
in \cite{I1} and it led to a new formula for the Hadamard's coefficients 
in terms of the $\tau$-function (or equivalently the wave and the adjoint 
wave functions) of the KdV hierarchy. This formula made transparent some of 
the basic properties of the Hadamard's coefficients such as the symmetry 
about the diagonal, or the connection between 
$H_n(x,x)$ and the KdV flows. We used this formula also in \cite{I2} to 
prove that heat kernel is finite (i.e. $H_n(x,y)=0$ for $n$ large enough) if 
and only if the potential $V(x)$ is among the rational solutions of 
the KdV hierarchy (for specific values of the time variables) studied in 
\cite{AM,AMcKM}. These operators appear also as solutions of the bispectral 
problem \cite{DG,Gr1}. For multivariable versions of the heat kernel and 
interesting connections with the bispectral problem and Huygens' principle 
see \cite{B,BV,CFV,CV}.

It is natural then to ask if one can use the soliton technology to study 
heat kernel expansions where very little or even nothing is known. 
The aim of the present paper is to show that there is a very close 
connection between the heat kernel expansion on the integers and the 
Toda lattice hierarchy. Some partial results in this direction were already 
established in \cite{Gr2,GI,H}, mainly in connection with the bispectral 
problem. Jointly with F.~A.~Gr\"unbaum \cite{GI} we studied the heat kernel 
for specific second-order difference operators which were introduced in 
\cite{HI2} as solutions of a 
difference-differential version of the bispectral problem. These second-order 
difference operators depend on free parameters and, appropriately normalized, 
they provide rational solutions of the Toda lattice hierarchy. The main result 
in \cite{GI} is that the fundamental solution corresponding to these operators 
can be written as a finite sum of Bessel functions of imaginary argument, 
which is a discrete analog of the finiteness property mentioned above for the 
continuous Schr\"odinger operator. The case when the operator $L$ corresponds 
to a soliton solution of the Toda lattice was considered in \cite{H}. Both 
papers rely on the fact that
the second-order operators belong to rank one commutative rings with 
spectral curves having specific singularities, which prevents their 
use for arbitrary $L$. The goal of this work is to derive a general formula 
for the analogs of Hadamard's coefficients in terms of the wave and the 
adjoint wave functions for the Toda lattice hierarchy and to present a few 
applications. In order to state the main results, let us first introduce some 
basic notations and define the analogs of Hadamard's coefficients 
in the discrete case.

We denote by $E$ the customary shift operator acting on 
functions $f(n)=f_n$ of a discrete variable $n\in\Z$ by
\begin{equation*}
E f(n) = f(n+1).
\end{equation*}
For a second-order difference operator $L$ of the form
\begin{equation}\label{1.1}
L=E+b_n\Id+a_nE^{-1},
\end{equation}
the fundamental solution (or the discrete heat kernel) is the solution 
$u(n,m;t)$ of the heat equation
\begin{equation}\label{1.2}
\frac{\pd u}{\pd t}=Lu,
\end{equation}
with initial condition 
\begin{equation}\label{1.3}
u(n,m;0)=\delta_{n,m}.
\end{equation}
The papers \cite{Gr2,GI,H} suggest to look for a solution of 
\eqref{1.2}-\eqref{1.3} of the form
\begin{equation}\label{1.4}
u(n,m;t)=\sum_{k=0}^{\infty}\alpha_k(n,m)I_{n-m-k}(2t),
\end{equation}
where $I_j(2t)$ is the Bessel function of imaginary argument. The coefficients 
$\alpha_k(n,m)$ are the analogs of the Hadamard's coefficients $H_k(x,y)$.
Plugging \eqref{1.4} in \eqref{1.2}, using the identity 
$$\pd_t I_k(2t)=I_{k+1}(2t)+I_{k-1}(2t)$$
and comparing the coefficients of $I_{n-m-k+1}(2t)$ we get
\begin{equation}\label{1.5}
\alpha_k(n,m)+\alpha_{k-2}(n,m)=\alpha_k(n+1,m)+b_n\alpha_{k-1}(n,m)
    +a_n\alpha_{k-2}(n-1,m),
\end{equation}
with the convention that $\alpha_i(n,m)=0$ if $i<0$. 
From \eqref{1.4} and using that $I_j(0)=\delta_{j,0}$ we obtain 
\begin{equation*}
u(n,m;0)=
\begin{cases} 0 & \text{if }n<m\\ 
\alpha_{n-m}(n,m)& \text{if }  n\geq m,
\end{cases}
\end{equation*}
which combined with \eqref{1.3} leads to
\begin{equation}\label{1.6}
\alpha_{n-m}(n,m)=\delta_{n,m} \text{ for every }n\geq m.
\end{equation}
It is clear that the system of equations \eqref{1.5}-\eqref{1.6} has a unique 
solution. Indeed, for $k=0$ equation \eqref{1.5} gives 
$\alpha_0(n,m)=\alpha_0(n+1,m)$, 
which essentially means that $\alpha_0(n,m)=\beta_m$ is independent of $n$. On 
the other hand, from equation  \eqref{1.6} we see that $\alpha_0(m,m)=1$, 
which shows that $\alpha_0(n,m)=1$. Similarly, for $k=1$ we obtain the 
system
\begin{align*}
&\alpha_1(n,m)=\alpha_1(n+1,m)+b_n\\
&\alpha_1(m+1,m)=0,
\end{align*}
which uniquely determines $\alpha_1(n,m)$. Next we can compute 
$\alpha_2(n,m)$, $\alpha_3(n,m)$, etc. We want to stress at this point that 
although the system \eqref{1.5}-\eqref{1.6} has a unique solution, the 
coefficients in the expansion 
\eqref{1.4} are uniquely determined from $u(n,m;t)$ only if $n\leq m$. 
Indeed, for $r\in\N_0$ we have 
\begin{equation}\label{1.7}
I_r(2t)=I_{-r}(2t)=\sum_{j=0}^{\infty}\frac{t^{r+2j}}{j!(r+j)!}.
\end{equation}
If $n\leq m$ then $\alpha_k(n,m)$ are uniquely determined by 
comparing the coefficients of the different powers of $t$ in \eqref{1.4}. 
However, for $n>m$ the expansion will contain both $I_r(2t)$ and 
$I_{-r}(2t)$ for $r=1,\dots, n-m$ and therefore only the sum of these 
2 coefficients is uniquely determined by $u(n,m;t)$. 
If we want to make the expansion unique, we can rearrange it so that the sum 
\eqref{1.4} contains only terms $I_j$ with $j\leq 0$. However, this would 
complicate significantly the recurrence relation \eqref{1.5} by introducing 
different cases when $n>m$ and $n\leq m$. Therefore we will keep the form 
\eqref{1.4} with $\alpha_k(n,m)$ uniquely determined by 
\eqref{1.5}-\eqref{1.6}.

Let me now briefly describe the main results in the paper. In the next 
section we collect some basic facts about the Toda lattice hierarchy. 
In \seref{se3} we derive simple formulas for the coefficients $\alpha_k(n,m)$
in terms of the wave and the adjoint wave function for the Toda lattice 
hierarchy (or equivalently the $\tau$-function). In other words, the soliton 
technology allows us to integrate explicitly equations 
\eqref{1.5}-\eqref{1.6} and write $\alpha_k(n,m)$ in closed form. As a first 
application of this formula, we show in \seref{se4} that $\alpha_k(n,n)$ and 
$\alpha_{k+1}(n,n-1)$ generate a hierarchy which is equivalent to the Toda 
lattice hierarchy (notice that we need two coefficients in order to generate 
two equations for $a_n$ and $b_n$). This provides a discrete analog of the 
connection between the Hadamard's coefficients $H_n(x,x)$ 
and the KdV hierarchy. It would be interesting to see if some of the 
proofs in the continuous case can be adapted to this situation. We note, 
however, that the situation in the discrete case is a little bit more 
complicated because the $k$-th vector field of the Toda lattice $\X_k$ is 
not equal to the vector field $\X_k'$ generated by $\alpha_k(n,n)$ 
and $\alpha_{k+1}(n,n-1)$, but we have
$\X_k'=\X_k+\text{a linear combination of }\X_{k-2j}$ with $2j<k$, see 
\thref{th4.4}. All this follows easily from the explicit formula for 
$\alpha_k(n,m)$, but clearly a direct approach should be carefully constructed 
in order to capture the lower-order vector fields that appear. 

Next we prove that the heat kernel can be written as 
\begin{equation}\label{1.8}
(1+p_1(n,m;t))I_{n-m}(2t)+p_2(n,m;t)I_{n-m-1}(2t),
\end{equation}
where $p_1(n,m;t)$ and $p_2(n,m;t)$ are polynomials in $t$ with coefficients 
depending on $n$ and $m$, such that $p_1(n,m;0)=p_2(n,m;0)=0$
if and only if the operator $L$ belongs to the family of bispectral operators 
constructed in \cite{HI2}. The ``if'' part is essentially the main result in 
\cite{GI} (see \seref{se6} for more details). 
In order to prove the ``only if'' part, we first discuss 
in \seref{se5} the operators constructed in \cite{HI2} and we show 
that they can be characterized by the vanishing of a specific linear 
combination of the Toda flows, after a particular point. 
The heart of the proof is to show that if this specific linear combination of 
the Toda flows vanishes after a particular point, then the operator $L$ 
belongs to a rank one commutative ring with spectral curve of the form 
$v^2=(u-2)^{2N_1+1}(u+2)^{2N_2+1}$, which leads to the operators introduced 
in \cite{HI2}. Using this result and the connection between the heat 
coefficients and the Toda flows, we establish the ``only if'' part in 
\ssref{ss6.2}.

The results in the last section deepen the mystery surrounding the bispectral 
problem and its connection with the heat equation. The finiteness property of 
the heat kernel provides one more reason to believe that the operators 
constructed in \cite{HI2} are the discrete analogs of the Adler-Moser 
operators \cite{AM}. It would be interesting to see if some of the recent 
developments \cite{Ho1,Ho2} in the purely continuous version of the bispectral 
problem can be extended to the operators in \cite{HI2}, or to the more 
general (higher-order) bispectral operators in \cite{HI1}, that we constructed 
in collaboration with L.~Haine, inspired by Wilson's work \cite{W1,W2}.

\section{The Toda lattice hierarchy}\label{se2}

In this section we briefly recall the construction of the Toda lattice 
hierarchy, its wave and adjoint wave functions. For more details we 
refer the reader to the paper \cite{UT}, or to the more recent 
contributions \cite{AvM,vM}.

Let us first introduce some notations, which will be used throughout the 
paper. We denote by $\Delta$ and $\nabla$ the customary forward 
and backward difference operators, acting on a function $f(n)=f_n$ as 
follows
\begin{align*}
&\Delta f(n)=f(n+1)-f(n)=(E-\Id)f(n)\\
&\nabla f(n)=f(n)-f(n-1)=(\Id-E^{-1})f(n).
\end{align*}
Sometimes, we will write $\Delta_n$ and $\nabla_n$ if the operators are 
applied to functions depending on several variables to indicate that 
$\Delta$ and $\nabla$ act in the variable $n$. 

When we work with pseudo-difference operators of the form \\
$M=\sum_{k=-\infty}^dc_k(n)E^k$ we will denote by $M_+=\sum_{k=0}^dc_k(n)E^k$ 
(resp. $M_-=\sum_{k=-\infty}^{-1}c_k(n)E^k$) 
the positive (resp. negative) difference part of $M$.

For a second-order difference operator $L=E+b_n\Id+a_nE^{-1}$, the Toda
lattice hierarchy is defined by the Lax equations
\begin{equation}\label{2.1}
\frac{\pd L}{\pd s_j}=[(L^j)_+,L],\text{ for }j=1,2,\dots.
\end{equation}
It is well-known that the vector fields $\X_j(L)=[(L^j)_+,L]$ commute with 
each other, that is, each of the equations \eqref{2.1} defines a symmetry 
of any other equation. For this reason, the family of equations \eqref{2.1} 
is called a hierarchy. The first equation, corresponding to $j=1$, is the 
well-known Toda lattice equation. 

We denote by $W_n=1+\sum_{k=1}^{\infty}\psi_k (n)E^{-k}$ the wave operator, 
i.e. the operator which conjugates $L$ to $E$
\begin{equation}\label{2.2}
L=W_nE W_n^{-1}.
\end{equation}
The flows \eqref{2.1} can be extended on $W_n$ by
\begin{equation}\label{2.3}
\frac{\pd W_n}{\pd s_j}=-(L^j)_{-}W_n.
\end{equation}

The wave function $\Psi_n(z;s)$ and the adjoint wave function $\Psi_n^*(z;s)$
are defined by the formulas
\begin{subequations}\label{2.4}
\begin{align}
\Psi_n(z;s)&=W_n z^n\exp\left(\sum_{j=1}^{\infty}s_jz^j\right)\nonumber\\
&=\left(1+\sum_{k=1}^{\infty}\frac{\psi_k (n;s)}{z^k}\right) z^n
\exp\left(\sum_{j=1}^{\infty}s_jz^j\right)\label{2.4a}\\
\intertext{and}
\Psi_n^*(z;s)
&=(W_{n-1}^{-1})^*z^{-n}\exp\left(-\sum_{j=1}^{\infty}t_jz^j \right)\nonumber\\
&=\left(1+\sum_{k=1}^{\infty}\frac{\psi_k^*(n;s)}{z^k}\right)z^{-n}
\exp\left(-\sum_{j=1}^{\infty}s_jz^j\right),\label{2.4b}
\end{align}
\end{subequations}
where $E^*=E^{-1}$. With these definitions we have 
\begin{equation}\label{2.5}
L\Psi_n(z;s)=z\Psi_n(z;s) \quad\text{ and }\quad 
\frac{\pd \Psi_n(z;s)}{\pd s_j}=(L^j)_+\Psi_n(z;s).
\end{equation}
We denote by $\Psib_n(z;s)$ and $\Psibs_n(z;s)$ the reduced wave function 
and the reduced adjoint wave function obtained from $\Psi_n(z;s)$ and 
 $\Psi^*_n(z;s)$, respectively, 
by omitting the exponential factors, i.e.
\begin{equation}\label{2.6}
\Psib_n(z;s)=1+\sum_{k=1}^{\infty}\frac{\psi_k (n;s)}{z^k} 
\quad\text{ and }\quad
\Psibs_n(z;s)=1+\sum_{k=1}^{\infty}\frac{\psi_k^* (n;s)}{z^k}.
\end{equation}
In order to simplify the notations, we will often omit the $s$-dependence, 
e.g. we will write simply $\Psi_n(z)$ instead of $\Psi_n(z;s)$, etc. Notice 
that the first equation in \eqref{2.5} is equivalent to the following equation 
for the reduced wave function
\begin{equation}\label{2.7}
z\Psib_{n+1}(z)+b_n\Psib_n(z)+a_n\frac{\Psib_{n-1}(z)}{z}=z\Psib_n(z).
\end{equation}

For series $\sum_kc_kz^k$ and for formal pseudo-difference operators
$\sum_kd_kE^k$ we denote 
\begin{equation*}
\res_z\left(\sum_kc_kz^k\right)=c_{-1}\quad\text{ and }\quad
\res_E\left(\sum_kd_kE^k\right)=d_{-1}.
\end{equation*}
It is easy to see that for pseudo-difference operators 
$P_n=\sum_{k}d_k(n)E^k$ and $Q_n=\sum_{k}c_k(n)E^{k}$
we have 
\begin{equation}\label{2.8}
\res_{z}\left((P_nz^n)(Q_{n-1}^*z^{-n})\right)=\res_{E}(P_nQ_n).
\end{equation}
From this equality it follows that $\Psi_n(z;s)$ and $\Psi_n^*(z;s)$
satisfy the following bilinear identities
\begin{equation}\label{2.9}
\res_{z}\left((\pd_{1}^{k_1}\pd_{2}^{k_2}\cdots\pd_{j}^{k_j}
\Psi_{n+k_0}(z;s))\Psi_{n}^*(z;s)\right)=0,
\end{equation}
where $k_0,k_1,\dots,k_j\in\N_0$ and $\pd_j=\frac{\pd}{\pd s_j}$. Indeed, 
from the second equation in \eqref{2.5} we see that it is enough to prove 
\eqref{2.9} for $k_1=k_2=\cdots=k_j=0$. Using now \eqref{2.8} we get 
\begin{align*}
&\res_{z}\left(\Psi_{n+k_0}(z;s))\Psi_{n}^*(z;s)\right)=
\res_{z}\left((E^{k_0}W_n z^n)((W^{-1}_{n-1})^*z^{-n})\right)\\
&\quad=\res_E(E^{k_0})=0,
\end{align*}
which establishes \eqref{2.9}. 

For example, for $k_0=0$ and $k_0=1$ with $k_j=0$ for $j\geq 1$, the 
identity \eqref{2.9} gives
\begin{subequations}\label{2.10}
\begin{align}
&\psi_1(n)+\psi_1^*(n)=0\label{2.10a}\\
&\psi_2(n+1)+\psi_1(n+1)\psi_1^*(n)+\psi_2^*(n)=0.\label{2.10b}
\end{align}
\end{subequations}

From \eqref{2.2} and \eqref{2.4} we can express $b_n$ and $a_n$ in terms 
of the coefficients of $\Psi_n(z;s)$ and $\Psi_n^*(z;s)$
\begin{subequations}\label{2.11}
\begin{align}
b_n&=\psi_1(n)+\psi_1^*(n+1)\label{2.11a}\\
a_n&=\psi_2(n)+\psi_1(n)\psi_1^*(n)+\psi_2(n).\label{2.11b}
\end{align}
\end{subequations}

\begin{Remark} \label{re2.1}
If we fix $n$ and take $k_0=0$, then the bilinear identities \eqref{2.9} 
show that $\Psi_n(z;s)z^{-n}$ and $\Psi_n^*(z;s)z^n$ are the wave and the 
adjoint wave functions of the Kadomtsev-Petviashvili (KP) hierarchy. Thus, by 
the classical theory (see \cite{DJKM}, \cite[Theorem 6.3.8, p.\,97]{Di}), 
there exits a $\tau$-function $\tau_n(s)$ such that 
\begin{equation*}
\Psib_n(z;s)=\frac{\tau_n(s-[z^{-1}])}{\tau_n(s)}
\text{ and }
\Psibs_n(z;s)=\frac{\tau_n(s+[z^{-1}])}{\tau_n(s)},
\end{equation*}
where $[z]=(z,z^2/2,z^3/3,\dots)$. This function plays a central role 
in the KP and the Toda lattice hierarchies. Since we are not going to 
make an explicit use of it in the present paper, we stop the discussion here. 
Notice that the last formulas allow us to express 
the coefficients of $\Psib_n(z;s)$ and $\Psibs_n(z;s)$ in terms of 
$\tau_n(s)$. In particular, this means that all formulas involving 
$\Psib_n(z;s)$ and $\Psibs_n(z;s)$ that follow can be rewritten as formulas
involving only $\tau_n(s)$.
\end{Remark}

\section{Explicit formulas for the heat coefficients $\alpha_k(n,m)$}
                                                                \label{se3}

In this section we show that, using the notations in the previous section, 
we can ``integrate'' equation \eqref{1.5} for all $k\in\N$ and obtain 
simple formulas for $\alpha_k(n,m)$ in terms of the wave and the adjoint 
wave functions of the Toda lattice hierarchy. Before we present and 
prove the general formula, we illustrate how the method works for 
$\alpha_1(n,m)$ and $\alpha_2(n,m)$. This will help the reader understand the 
nature of these formulas and the importance of the bilinear identity 
\eqref{2.9}.

Plugging $k=1$ in \eqref{1.5} and using \eqref{2.11} we get
$$\alpha_1(n,m)=\alpha_1(n+1,m)+\psi_1(n)+\psi_1^*(n+1).$$
Using \eqref{2.10a} we can rewrite the last equation as
$$\alpha_1(n,m)-\psi_1(n)=\alpha_1(n+1,m)-\psi_1(n+1),$$
which shows that $\alpha_1(n,m)-\psi_1(n)=\beta_1(m)$ is a function 
independent of $n$. From \eqref{1.6} we see that $\alpha_1(m+1,m)=0$, which 
gives $\beta_1(m)=-\psi(m+1)=\psi_1^*(m+1)$. Thus we have 
\begin{equation}\label{3.1}
\alpha_1(n,m)=\psi_1(n)+\psi_1^*(m+1)
=\res_{z} \left[\Psib_n(z)\Psibs_{m+1}(z)\right].
\end{equation}
Similarly, for $k=2$ we obtain the following equation for $\alpha_2(n,m)$
\begin{align}
\alpha_2(n,m)+1&=\alpha_2(n+1,m)
+[\psi_1(n)+\psi_1^*(n+1)][\psi_1(n)+\psi_1^*(n+1)]\nonumber\\
&\quad+\psi_2(n)+\psi_1(n)\psi_1^*(n)+\psi_2^*(n).\label{3.2}
\end{align}
Using now both equations in \eqref{2.10} one can check that \eqref{3.2} 
is equivalent to the equation
\begin{equation*}
\Delta_n\left(\alpha_2(n,m)-\psi_2(n)-\psi_1(n)\psi_1^*(m+1)-n\right)=0,
\end{equation*}
which means that 
$\alpha_2(n,m)=\psi_2(n)+\psi_1(n)\psi_1^*(m+1)+n+\beta_2(m)$, where 
$\beta_2(m)$ depends only on $m$. From \eqref{1.6} we see that 
$\alpha_2(m+2,m)=0$ leading to $\beta_2(m)=-\psi_2(m+2)
-\psi_1(m+2)\psi_1^*(m+1)-m-2=\psi_2^*(m+1)-m-2$, where in the last equality 
we used again \eqref{2.10b}. Thus we obtain
\begin{align}
\alpha_2(n,m)&=\psi_2(n)+\psi_1(n)\psi_1^*(m+1)+\psi_2^*(m+1)+(n-m-2)
        \nonumber\\
&=\res_{z} \left[(z^2+n-m-2)\frac{\Psib_n(z)\Psibs_{m+1}(z)}{z}\right].
        \label{3.3}
\end{align}
The point is that this process can be successfully iterated providing simple 
formulas for $\alpha_k(n,m)$ for every $k\in\N$. Below we introduce the 
necessary functions which are needed to state and prove the general formula, 
extending \eqref{3.1} and \eqref{3.3}.

Let us define monic polynomials $Q^{\beta}_k(z)$ by the formula
\begin{equation}\label{3.4}
Q^{\beta}_k(z)=z^k+(\beta-2k)\sum_{j=0}^{k-1}\binom{\beta-2j-1}{k-j-1}
\frac{z^j}{k-j}.
\end{equation}
From \eqref{3.4} it is easy to see that 
\begin{align}
&Q^{\beta}_k(z)-zQ^{\beta-2}_{k-1}(z)=\frac{\beta-2k}{k}\binom{\beta-1}{k-1}
\label{3.5}\\
\intertext{and}
&Q^{\beta}_{k}(z)+Q^{\beta}_{k-1}(z)-zQ^{\beta-1}_{k-1}(z)
=\frac{\beta-2k+1}{k}\binom{\beta}{k-1}.\label{3.6}
\end{align}
Let us denote 
\begin{equation}\label{3.7}
g_k(n,m;z)=
\begin{cases} Q^{n-m}_{k/2}(z^2) & \text{if $k$ is even}\\ 
zQ^{n-m-1}_{(k-1)/2}(z^2)& \text{if $k$ is odd.}
\end{cases}
\end{equation}
It is clear that $g_k(n,m;z)$ is a monic polynomial in $z$ of degree $k$, and 
it is an even polynomial when $k$ is even and an odd polynomial when $k$ is 
odd. From the defining relation \eqref{3.7} and equation \eqref{3.5} one sees
immediately that
\begin{subequations}\label{3.8}
\begin{align}
&g_k(n,m;z)-zg_{k-1}(n-1,m;z)=0 \qquad\text{ if $k$ is odd}\label{3.8a}\\
\intertext{and}
&g_k(n,m;z)-zg_{k-1}(n-1,m;z)=
g_k(n,m;z)-z^2g_{k-2}(n-2,m;z)\nonumber\\
&\qquad=\frac{2(n-m-k)}{k}\binom{n-m-1}{k/2-1}\qquad
\text{ if $k$ is even.}\label{3.8b}
\end{align}
\end{subequations}
Now we are ready to formulate the main result in this section which gives 
an explicit formula for $\alpha_k(n,m)$ in terms of $\Psib_n(z)$ and 
$\Psibs_n(z)$.
\begin{Theorem} \label{th3.1}
The coefficients $\alpha_k(n,m)$ in the expansion \eqref{1.4} of the 
fundamental solution of the discrete heat equation \eqref{1.2} can be 
expressed in terms of the reduced wave function $\Psib_n(z)$ and the 
reduced adjoint wave function $\Psibs_n(z)$ as follows 
\begin{equation}\label{3.9}
\alpha_k(n,m)=
\res_{z} \left[g_k(n,m;z)\frac{\Psib_n(z)\Psibs_{m+1}(z)}{z}\right],
\end{equation}
where $g_k(n,m;z)$ is the polynomial defined by \eqref{3.4} and \eqref{3.7}.
\end{Theorem}

\begin{proof} We need to check that the functions defined by 
formula \eqref{3.9} satisfy the difference equation \eqref{1.5} and the 
initial conditions \eqref{1.6}. Let us first establish that equation 
\eqref{1.5} holds. We denote by RHS (resp. LHS) the right-hand side 
(resp. the left-hand side) of equation \eqref{1.5}.
Plugging \eqref{3.9} in \eqref{1.5} leads to 
\begin{align*}
\RHS&=\alpha_k(n+1,m)+b_n\alpha_{k-1}(n,m)+a_n\alpha_{k-2}(n-1,m) \\
&= \res_z\Bigg[g_k(n+1,m;z)\frac{\Psib_{n+1}(z)\Psibs_{m+1}(z)}{z}
+b_n g_{k-1}(n,m;z)\frac{\Psib_{n}(z)\Psibs_{m+1}(z)}{z}\\
&\qquad+a_n g_{k-2}(n-1,m;z)\frac{\Psib_{n-1}(z)\Psibs_{m+1}(z)}{z}\Bigg]\\
&=\res_z\Bigg[\left((z^2-zb_n)g_{k-2}(n-1,m;z)+b_ng_{k-1}(n,m;z)\right)
\frac{\Psib_{n}(z)\Psibs_{m+1}(z)}{z} \\
&\qquad +\left(g_{k}(n+1,m;z)-z^2g_{k-2}(n-1,m;z)\right)
\frac{\Psib_{n+1}(z)\Psibs_{m+1}(z)}{z}\Bigg],
\end{align*}
where in the last equality we eliminated $\Psib_{n-1}(z)$ using \eqref{2.7}.

Now we consider two cases depending on whether $k$ is even or odd.\\

{\em Case 1.\/} Let $k$ be even (hence $k-1$ is odd). Using \eqref{3.8} 
we see that
$$(z^2-zb_n)g_{k-2}(n-1,m;z)+b_ng_{k-1}(n,m;z)
=z^2g_{k-2}(n-1,m;z)$$
and 
$$g_{k}(n+1,m;z)-z^2g_{k-2}(n-1,m;z)=
\frac{2(n+1-m-k)}{k}\binom{n-m}{k/2-1}$$
which lead to the following expressions for RHS and LHS:
\begin{align*}
\RHS&=\res_z \bigg[z^2g_{k-2}(n-1,m;z)\frac{\Psib_n(z)\Psibs_{m+1}(z)}{z}\bigg]
                        \\
&\qquad+\frac{2(n+1-m-k)}{k}\binom{n-m}{k/2-1}\\
\intertext{and}
\LHS& = \res_z\bigg[(g_k(n,m;z)+g_{k-2}(n,m;z))
\frac{\Psib_n(z)\Psibs_{m+1}(z)}{z}\bigg].
\end{align*}
Using now equation \eqref{3.6} we get
\begin{align*}
&g_{k}(n,m;z)+g_{k-2}(n,m;z)-z^2g_{k-2}(n-1,m;z) \\
&\qquad =
Q^{n-m}_{k/2}(z^2)+Q^{n-m}_{k/2-1}(z^2)-z^2Q^{n-m-1}_{k/2-1}(z^2)\\
&\qquad = \frac{2(n+1-m-k)}{k}\binom{n-m}{k/2-1},
\end{align*}
which shows that $\RHS=\LHS$ and establishes \eqref{1.5}.\\

{\em Case 2.\/} Let $k$ be odd (hence $k-1$ is even). 

Using again \eqref{3.8} we obtain 
\begin{align*}
&g_k(n+1,m;z)-z^2g_{k-2}(n-1,m;z)\\
&\qquad =z(g_{k-1}(n,m;z)-z^2g_{k-3}(n-2,m;z))\\
&\qquad=
\frac{2(n-m-k+1)}{k-1}\binom{n-m-1}{(k-1)/2-1}z
\end{align*}
and 
\begin{align*}
&(z^2-zb_n)g_{k-2}(n-1,m;z)+b_ng_{k-1}(n,m;z) \\
&\qquad =
(z^2-zb_n)zg_{k-3}(n-2,m;z)+b_ng_{k-1}(n,m;z)\\
&\qquad =
b_n(g_{k-1}(n,m;z)-z^2g_{k-3}(n-2,m;z))+z^3g_{k-3}(n-2,m;z)\\
&\qquad =
b_n \frac{2(n-m-k+1)}{k-1}\binom{n-m-1}{(k-1)/2-1} 
+z^3g_{k-3}(n-2,m;z).
\end{align*}
Thus
\begin{align*}
\RHS & = \res_z\bigg[\frac{2(n-m-k+1)}{k-1}\binom{n-m-1}{(k-1)/2-1}
(z\Psib_{n+1}(z)+b_n\Psib_n(z))\frac{\Psibs_{m+1}(z)}{z} \\
&\qquad +z^3g_{k-3}(n-2,m;z)\frac{\Psib_n(z)\Psibs_{m+1}(z)}{z}\bigg]\\
& = \res_z\bigg[
\bigg(\frac{2(n-m-k+1)}{k-1}\binom{n-m-1}{(k-1)/2-1}+z^2g_{k-3}(n-2,m;z)\bigg)
\\
&\qquad\qquad\times \Psib_n(z)\Psibs_{m+1}(z)\bigg],
\end{align*}
where in the last equality we used \eqref{2.7} to eliminate 
$z\Psib_{n+1}(z)+b_n\Psib_n(z)$.
On the other hand we have
$$
\LHS=\res_z\big[(g_{k-1}(n-1,m;z)+g_{k-3}(n-1,m;z))
         \Psib_n(z)\Psibs_{m+1}(z)\big],$$
and applying the definition \eqref{3.7} and equation \eqref{3.6} we get
\begin{align*}
&g_{k-1}(n-1,m;z)+g_{k-3}(n-1,m;z)-z^2g_{k-3}(n-2,m;z)\\
&\qquad =
Q^{n-m-1}_{(k-1)/2}(z^2)+Q^{n-m-1}_{(k-1)/2-1}(z^2)
-z^2Q^{n-m-2}_{(k-1)/2-1}(z^2)\\
&\qquad =\frac{2(n-m-k+1)}{k-1}\binom{n-m-1}{(k-1)/2-1},
\end{align*}
which shows that $\RHS=\LHS$ and completes the proof of the difference 
equation \eqref{1.5}.

It remains to check that the initial condition \eqref{1.6} holds, i.e. we 
need to show that for every $k\in\N$,  $\alpha_{k}(m+k,k)=0$. 
Using the defining relations \eqref{3.4} and \eqref{3.7} one can easily see 
that 
$$g_k(m+k,m;z)=z^k.$$
Thus \eqref{3.9} gives
$$\alpha_{k}(m+k,m)=
\res_z\Big[z^{k}\frac{\Psib_{m+k}(z)\Psibs_{m+1}(z)}{z}\Big]
=\res_z[\Psi_{m+k}(z)\Psi^*_{m+1}(z)]=0,$$
where in the last equality we used the bilinear identity \eqref{2.9}.
\end{proof}

\section{Generating the Toda flows with the heat coefficients}\label{se4}

As a first application of formula \eqref{3.9} we prove in this section 
that \\$\Delta_n\alpha_{k+1}(n,n-1)$ and $a_n\nabla_n\alpha_k(n,n)$ generate 
the Toda lattice hierarchy. This is a discrete analog of the remarkable 
connection 
between the restriction of the Hadamard's coefficients on the diagonal and 
the Korteweg-de Vries hierarchy. Before we prove this, we 
establish some auxiliary facts.

\begin{Proposition}\label{pr4.1}
Let us denote
\begin{subequations}\label{4.1}
\begin{align}
&\fr_k(n)=\res_{z}\left(z^{k}\Psib_{n}(z)\Psibs_{n}(z)\right)\label{4.1a}\\
&\fl_k(n)=\res_{z}\left(z^{k-1}\Psib_{n-1}(z)\Psibs_{n}(z)\right).\label{4.1b}
\end{align}
\end{subequations}
Then the Toda lattice hierarchy \eqref{2.1} is equivalent to the equations
\begin{subequations}\label{4.2}
\begin{align}
&\frac{\pd b_n}{\pd s_k}=\fr_k(n+1)-\fr_k(n)\label{4.2a}\\
&\frac{\pd a_n}{\pd s_k}=a_n(\fl_k(n+1)-\fl_k(n)).\label{4.2b}
\end{align}
\end{subequations}
\end{Proposition}
\begin{proof}
Notice that equation \eqref{2.4b} implies
\begin{equation*}
W_n^{-1}=1+\sum_{j=1}^\infty E^{-j}\cdot \psi_j^*(n+1),
\end{equation*}
and therefore, using \eqref{2.2} and \eqref{2.4a}, we get
\begin{equation*}
L^k=W_nE^kW_n^{-1}=\sum_{m=0}^{\infty}\left[
\sum_{j=0}^m\psi_{m-j}(n)\psi^*_j(n+1+k-m)\right]E^{k-m},
\end{equation*}
where $\psi_0(n)=\psi_0^*(n)=1$.
From the above equation it is clear that the coefficients of $E^0$ and 
$E^{-1}$ in the operator $L^k$ are $\fl_k(n+1)$ and $\fr_k(n)$ respectively. 
This shows that 
\begin{align*}
[(L^k)_+,L]&=[L,(L^{k})_-]=[E+b_n\Id+a_nE^{-1},\fr_k(n)E^{-1}+O(E^{-2})]\\
&=(\fr_k(n+1)-\fr_{k}(n))\Id+O(E^{-1}),
\end{align*}
which gives \eqref{4.2a}. Similar computation shows that the coefficient of 
$E^{-1}$ in $[(L^k)_+,L]$ is $a_n(\fl_k(n+1)-\fl_k(n))$, completing the proof.
\end{proof}

\begin{Lemma} \label{le4.2}
Let $k\in\N_0$ and $j\in\Z$ be fixed. Then \\
$\alpha_k(n,n+j)\in
\C[a_{n},b_{n},a_{n\pm1},b_{n\pm 1},a_{n\pm2},b_{n\pm 2},\dots]$. In other 
words, \\ $\alpha_k(n,n+j)$ is a polynomial of finitely many of 
$\{a_{n},b_{n},a_{n\pm1},b_{n\pm 1},\dots\}$.
\end{Lemma}

\begin{proof} The proof can be easily obtained by induction on $k$ and 
$|k+j|$, using \eqref{1.5}-\eqref{1.6}.
\end{proof}

\begin{Example} \label{ex4.3}
We list the values of the first few heat coefficients in a 
neighborhood of the diagonal $n=m$ 
\begin{align*} 
&\alpha_1(n,n)=b_n,\\ 
&\alpha_1(n,n-1)=0, \\
&\alpha_1(n,n+1)=b_n+b_{n+1}\\
&\alpha_2(n,n)=a_{n+1}+a_{n}+b_{n}^2-2,\\ 
&\alpha_2(n,n-1)=a_{n}-1, \\
&\alpha_2(n,n+1)=a_{n}+a_{n+1}+a_{n+2}+b_{n}^2+b_{n+1}^2+b_nb_{n+1}-3.
\end{align*}
\end{Example}

The main result in this section is the following theorem.
\begin{Theorem} \label{th4.4}
The system of differential-difference equations
\begin{subequations}\label{4.3}
\begin{align}
&\frac{\pd b_n}{\pd s'_k}=\alpha_{k+1}(n+1,n)-\alpha_{k+1}(n,n-1)
=\Delta_n\alpha_{k+1}(n,n-1)\label{4.3a}\\
&\frac{\pd a_n}{\pd s'_k}=a_n(\alpha_{k}(n,n)-\alpha_{k}(n-1,n-1))
=a_n\nabla_n\alpha_{k}(n,n),\label{4.3b}
\end{align}
\end{subequations}
where $k\in\N$ forms a hierarchy. Moreover, if denote by $\X_k$ and $\X'_k$ 
the vector fields corresponding to the flows $\pd/\pd s_k$ and 
$\pd/\pd s'_k$ given by \eqref{2.1} and \eqref{4.3} respectively, then 
\begin{equation}\label{4.4}
\X'_k=k\sum_{i=0}^{\lfloor(k-1)/2\rfloor}
\frac{(-1)^i}{k-2i}\binom{k-i-1}{i}\X_{k-2i},
\end{equation}
where $\lfloor x\rfloor$ denotes the greatest integer less than or equal to 
$x$.
\end{Theorem}

\begin{Remark} Notice that according to \leref{le4.2}, for every $k\in\N$ 
the right-hand sides of \eqref{4.3} are polynomials of finitely many of 
 $\{a_{n+j},b_{n+j}\}_{j\in\Z}$, i.e. 
\eqref{4.3} is a well-defined system of differential equations for 
$a_n$ and $b_n$.
The above theorem essentially says that the hierarchy of 
equations \eqref{4.3} is equivalent to the Toda lattice hierarchy \eqref{2.1} 
modulo a simple 
(linear) change of variables given by \eqref{4.4}. Below we write also 
the explicit linear combination that gives $\X_k$ in terms of $\{\X'_j\}$, 
which will be needed later. First we show that for every $k\in\N$ and 
$m\in\N_0$ satisfying $m\leq\lfloor(k-1)/2\rfloor$ the following identity
\begin{equation}\label{4.5}
k\sum_{i=0}^m\frac{(-1)^i}{k-2i}\binom{k-i-1}{i}\binom{k-2i}{m-i}=\delta_{m,0}
\end{equation}
holds. Indeed, if $m=0$ then \eqref{4.5} is obvious. For $m\geq 1$ we can 
rewrite the left-hand side of \eqref{4.5} as
\begin{equation*}
\frac{k}{m}\sum_{i=0}^m(-1)^i\binom{k-1-i}{m-1}\binom{m}{i}
=\frac{k}{m}\binom{k-1-m}{k-m}=0.
\end{equation*}
In the first equality we used the well-known binomial identity
\begin{equation}\label{4.6}
\sum_{i=0}^m(-1)^i\binom{k-i}{r}\binom{m}{i}=\binom{k-m}{k-r},
\end{equation}
which can be easily proved by applying the principle of inclusion and 
exclusion to the following problem: In how many ways one can select $r$
of given $k$ distinct objects, so that each selection includes some particular 
$m$ of the $k$ objects.

Combining \eqref{4.4} and \eqref{4.5} one can deduce that
\begin{equation}\label{4.7}
\X_k=\sum_{j=0}^{\lfloor(k-1)/2\rfloor}\binom{k}{j}\X'_{k-2j}.
\end{equation}
\end{Remark}

\begin{proof}[Proof of \thref{th4.4}]
It is enough to prove \eqref{4.4}, because this formula and the fact that 
$\{\X_k\}$ commute will imply that $\{\X_k'\}$ commute, i.e. the equations 
\eqref{4.3} form a hierarchy.

Using \eqref{3.4} and \eqref{3.7} one can easily check that
\begin{equation}\label{4.8}
g_{k+1}(n,n-1;z)=z^{k+1}+k\sum_{i=1}^{\lfloor(k+1)/2\rfloor}\frac{(-1)^i}{i}
\binom{k-i-1}{i-1}z^{k+1-2i},
\end{equation}
which combined with \eqref{3.9} and \eqref{4.1a} gives 
\begin{equation}\label{4.9}
\alpha_{k+1}(n,n-1)=\fr_{k}(n)+k
\sum_{i=1}^{\lfloor(k+1)/2\rfloor}\frac{(-1)^i}{i}
\binom{k-i-1}{i-1}\fr_{k-2i}(n).
\end{equation}
On the other hand, from equation \eqref{3.8} it follows that 
\begin{equation*}
g_{k+1}(n,n-1;z)=zg_k(n-1,n-1;z)-\delta_{k,1},
\end{equation*}
which shows that
\begin{equation}\label{4.10}
g_{k}(n-1,n-1;z)=z^{k}+k\sum_{i=1}^{\lfloor(k+1)/2\rfloor}\frac{(-1)^i}{i}
\binom{k-i-1}{i-1}z^{k-2i}+\frac{\delta_{k,1}}{z},
\end{equation}
and therefore, using \eqref{3.9} and \eqref{4.1b} we get
\begin{equation}\label{4.11}
\alpha_{k}(n-1,n-1)=\fl_{k}(n)
+k\sum_{i=1}^{\lfloor(k+1)/2\rfloor}\frac{(-1)^i}{i}
\binom{k-i-1}{i-1}\fl_{k-2i}(n).
\end{equation}
The proof now follows from \eqref{4.9}, \eqref{4.11}, \prref{pr4.1} and 
the fact that $\fr_0(n)=0$, $\fr_{-1}(n)=1$, $\fl_0(n)=1$, $\fl_{-1}(n)=0$ are 
independent of $n$.
\end{proof}

\section{Darboux transformations from $L_0=E+E^{-1}$ at the end points of the
spectrum}\label{se5}

In this section we focus on certain second-order difference operators 
$L_{N_1,N_2}$, which were introduced in \cite{HI2} in connections with a 
difference-differential version of the bispectral problem \cite{DG}. 
These operators can be defined by successive Darboux transformations 
from the operator $L_0=E+E^{-1}$ at the end points $\pm 2$ of the spectrum.  
Recall (see \cite{MS}) that the Darboux transformation of a second-order 
operator $L$ at point $c_0$ consists of factorizing $L-c_0\Id$ as a product 
of first-order operators and producing a new operator $\hat{L}$ by exchanging 
the factors, i.e. if we write $L-c_0\Id=\cP\cQ$, then $\hat{L}$ is defined by 
$\hat{L}-c_0\Id=\cQ\cP$. The main result in this section is a characterization 
of these operators in terms of the vector fields of the Toda lattice 
hierarchy. This is needed in the next section where we prove that the heat 
kernel expansion for these operators can be written as a sum of only two 
Bessel functions with polynomial coefficients (in the time variable $t$), 
and that this property completely characterizes the operators $L_{N_1,N_2}$.

\subsection{Constructing the operators $L_{N_1,N_2}$}\label{ss5.1}

The operators $L_{N_1,N_2}$ are obtained by the following sequence of the 
Darboux transformations
\begin{align}
&L_{0}-2\,\Id=\cP_{0}\cQ_{0}\curvearrowright 
L_{1,0}-2\,\Id=\cQ_{0}\cP_{0}=\cP_{1}\cQ_{1}
\curvearrowright\cdots \nonumber \\
&\qquad \curvearrowright L_{N_1,0}-2\,\Id=\cQ_{N_1-1}\cP_{N_1-1}   \nonumber\\
&L_{N_1,0}+2\,\Id=\cP_{N_1}\cQ_{N_1} \curvearrowright
L_{N_1,1}+2\,\Id=\cQ_{N_1}\cP_{N_1}= \cP_{N_1+1}\cQ_{N_1+1}
\curvearrowright\cdots
                                                                  \nonumber\\
&\qquad\curvearrowright L_{N_1,N_2}+2\,\Id=\cQ_{N_1+N_2-1}\cP_{N_1+N_2-1}.
\label{5.1}
\end{align}
At each step, the factorization of the operator $L_{i_1,i_2}\pm2\Id$ depends 
on one free parameter. Thus, the operator $L_{N_1,N_2}$ will depend on 
$N_1+N_2$ free parameters. The operator $L_{N_1,N_2}$ belongs to a rank one 
commutative ring of difference operators, i.e. we can apply the 
correspondence between commutative rings of difference operators 
and algebraic curves developed in the papers \cite{Kr,Mum,vMM}. 
We first sketch the main steps of this construction with an emphasize on the 
operators obtained by the Darboux process \eqref{5.1} and refer the reader 
to \cite{HI1,HI2} for more details.

Following \cite{Mum}, we 
call a difference operator $M=\sum_{k=K_-}^{K_+}\mu_k(n)E^k$ properly bordered 
if $\mu_{K_-}(n)\neq 0$ and $\mu_{K_+}(n)\neq 0$ for all $n\in\Z$; the 
interval $[K_-,K_+]$ is the support of $M$. A commutative ring $\cA$ of 
difference operators is called rank one, if it contains two properly bordered 
difference operators 
$M'$ and $M''$ with supports $[K'_-,K'_+]$ and $[K''_-,K''_+]$ such that 
$\gcd(K'_-,K''_-)=1$, $\gcd(K'_+,K''_+)=1$ and $K'_-K''_+<K'_+K''_-$. In that 
case, $\Spec(\cA)$ is an irreducible complex affine curve that completes 
by adding two nonsingular points $Q^{\pm}_{\infty}$ at infinity.

Starting with a properly bordered second-order difference operator 
$L=E+b_n\Id+a_nE^{-1}$ we denote by $\cA_L$ the ring of all difference 
operators commuting with $L$, i.e.
$$\cA_L=\left\{M=\sum_{k=K_-}^{K_+}\mu_k(n)E^k:[M,L]=0\right\}.$$
One can show that $\cA_L$ is in fact a commutative ring consisting of 
properly bordered difference operators and it is a rank one ring if and only 
$\cA_L$ contains an operator which is not a polynomial of $L$. For every 
$M\in\cA_L$ the operators $L$ and $M$ satisfy an algebraic relations of the 
form
\begin{equation*}
M^2=MT_1(L)+T_2(L),
\end{equation*}
where $T_1$ and $T_2$ are some polynomials. This equation defines an affine 
curve $f_{L,M}(u,v)=v^2-v T_1(u)-T_2(u)=0$. It is easy to see that 
for every $M\in\cA_L$ there exists a unique polynomial $g_M$ such that 
$g_M(0)=0$ and the 
operator $M-g_M(L)$ contains only nonnegative powers of $E$, i.e. 
$M=g_M(L)+\sum_{j=0}^{\nu(M)}c_k(n)E^{k}$. From this, it follows that 
the ring $\cA_L$ is generated by two operators $\{L,M\}$ where $M$ is chosen 
so that $\nu(M)>0$ is minimal. The spectral curve $\Spec(\cA_L)$ is 
$f_{L,M}(u,v)=0$ and the complete curve is 
$X_L=\Spec(\cA_L)\cup\{Q^+_\infty,Q^-_\infty\}$.

The Baker function $\Psi_n$ for $\cA_L$ is the unique (up to a factor 
independent of $n$) eigenfunction for the operators from $\cA_L$. If we 
denote by $A_L$ the ring of functions meromorphic on 
$X$ with poles only at $Q^{\pm}_{\infty}$, then for every $M\in\cA_L$ we have
$$M\Psi_n(P)=h_M(P)\Psi_n(P), \text{ where } h_M(P)\in A_L.$$
Moreover, if the support of $M$ is $[K_-,K_+]$ then 
$h_M(P)$ has poles of orders $K_+$ and $K_-$ at $Q^+_\infty$ and 
$Q^-_\infty$, respectively.

Let us denote by $\cA_{N_1,N_2}=\cA_{L_{N_1,N_2}}$ the ring of all difference 
operators, commuting with $L_{N_1,N_2}$. In \cite{HI2} it was shown that 
$\cA_{N_1,N_2}$ is a rank one commutative ring of difference operators, 
which is isomorphic to the ring 
\begin{equation*}
A_{N_1,N_2}=A_{L_{N_1,N_2}}=\C[x+x^{-1},f_{N_1,N_2}(x)]\subset\C[x,x^{-1}], 
\end{equation*}
where 
\begin{equation}\label{5.2}
f_{N_1,N_2}(x)=\frac{(x-1)^{2N_1+1}(x+1)^{2N_2+1}}{x^{N_1+N_2+1}}.
\end{equation}
In other words, if $\Psi_n$ is the Baker function for $\cA_{N_1,N_2}$ then 
\begin{subequations}\label{5.3}
\begin{align}
L_{N_1,N_2}\Psi_n(x)&=(x+x^{-1})\Psi_n(x)\label{5.3a}\\
M_{N_1,N_2}\Psi_n(x)&=f_{N_1,N_2}(x)\Psi_n(x),\label{5.3b}
\end{align}
\end{subequations}
for some $M_{N_1,N_2}\in\cA_{N_1,N_2}$ and the ring $\cA_{N_1,N_2}$ is 
generated by $L_{N_1,N_2}$ and $M_{N_1,N_2}$.
The spectral curve is given by the equation
\begin{equation}\label{5.4}
\Spec(\cA_{N_1,N_2}):\quad v^2=(u-2)^{2N_1+1}(u+2)^{2N_2+1}.
\end{equation}
Clearly, $\Spec(\cA_{N_1,N_2})$ is rational and has a cusp at $u=2$ (resp. 
$u=-2$) when $N_1>0$ (resp. $N_2>0$). 
Another property, that we will need later, is that for every $j\in\Z$ we have
\begin{equation}\label{5.5}
x^jf_{N_1,N_2}(x)\in A_{N_1,N_2},
\end{equation}
see \cite[the proof of Theorem 4.2.]{HI2} 
(note that $x$ here corresponds to $z+1$ in \cite{HI2}).

Conversely, let $L=E+b_n\Id+a_nE^{-1}$ be a properly bordered second-order 
difference operator whose spectral curve is given by equation \eqref{5.4}. 
From the 
correspondence in \cite{Mum} we can conclude that $\cA_L$ is obtained 
from $\cA_{N_1,N_2}$ for a specific choice of the free parameters in the 
Darboux steps, up to a conjugation by a nonzero function $g_n$, i.e.
\begin{equation}\label{5.6}
\cA_L=\{g_n^{-1}Mg_n: M\in\cA_{N_1,N_2}\}.
\end{equation}
If $L$ is nonconstant (i.e. at least one of the functions $a_n$ and $b_n$ is 
not a constant), then 
it is easy to see that all second-order operators operators in $\cA_L$ 
with support $[-1,1]$ must have the form $\gamma_1 L+\gamma_2\Id$ for some 
constants $\gamma_1,\gamma_2$. From this, it follows that the only 
possible functions $g_n$ in \eqref{5.6} are $g_n=c^n$, and we must have
\begin{equation}\label{5.7}
L=c^{-n-1}L_{N_1,N_2}c^{n}+d\,\Id,
\end{equation}
for some constants $c\neq 0$ and $d$. 

One possible way to eliminate this freedom and characterize precisely the 
operators $L_{N_1,N_2}$ is to consider properly bordered second-order 
difference operators of the form  $L=E+b_n\Id+a_nE^{-1}$ with spectral 
curve  given in \eqref{5.4} and coefficients 
satisfying
\begin{equation}\label{5.8}
\lim_{n\rightarrow\infty}b_n=0, \qquad \lim_{n\rightarrow\infty}a_n=1.
\end{equation}
Indeed, if $L=L_{N_1,N_2}$ then the coefficients $b_n$, $a_n$ can be 
computed from the formulas
$$b_n=\frac{\pd}{\pd s_1}\log\frac{\bar{\tau}_{n+1}(s)}{\bar{\tau}_{n}(s)},
\quad 
a_n=\frac{\bar{\tau}_{n+1}(s)\bar{\tau}_{n-1}(s)}{\bar{\tau}_{n}(s)^2},$$
where $\bar{\tau}_n(s)$ is a polynomial in $n$, which makes \eqref{5.8} 
obvious. We note, however, that $\bar{\tau}_n(s)$ 
in the last formulas differs from the $\tau$-function introduced in 
\seref{se2} (due to the different approach in the papers \cite{HI1,HI2}).
Conversely, if $L$ is nonconstant coefficient operator which belongs to a rank 
one commutative ring of difference operators whose spectral curve is given by 
\eqref{5.4}, then \eqref{5.7} must hold which combined 
with \eqref{5.8} shows that $c=\pm 1$ and $d=0$. 
Conjugating $L_{N_1,N_2}$ by $(-1)^n$ essentially exchanges the roles of 
$+2$ and $-2$ in \eqref{5.1} (or, equivalently, the roles of $N_1$ and 
$N_2$). Thus, we have $L=L_{N_1,N_2}$ or $L=L_{N_2,N_1}$, completing the 
proof in this case. Finally, if $a_n$ and $b_n$ are constants, then 
\eqref{5.8} implies that $L=L_0$.

Next, we establish several new facts needed for the 
characterization of the operators $L_{N_1,N_2}$ in terms of the Toda vector 
fields proved at the end of this section.

\begin{Lemma} \label{le5.1}
Let $A$ be a ring of Laurent polynomials in $x$ such that \\
$A_{N_1,N_2}\subset A$ for some $N_1,N_2\in \N_0$. If $A$ 
contains a polynomial $p(x)$ such that $0<\deg(p(x))\leq N_1+N_2$ then one of 
the following must hold
\begin{itemize}
\item[(i)] $N_1\geq 1$ and $f_{N_1-1,N_2}(x)\in A$; 
\item[(ii)] $N_2\geq 1$ and $f_{N_1,N_2-1}(x)\in A$.
\end{itemize}
\end{Lemma}

\begin{proof}
Assume that $p(x)\in A\cap\C[x]$ is such that $0<\deg(p(x))\leq N_1+N_2$.
Since $x+x^{-1}\in A$, it is clear that $p(x)+p(1/x)\in A$, which combined 
with $p(x)\in A$ shows that $f(x)=p(x)-p(1/x)\in A$. Notice that 
$f(1/x)=-f(x)$ and therefore we can write $f(x)$ in the form 
\begin{equation}\label{5.9}
f(x)=\left(x-\frac{1}{x}\right)^{2l+1}\ft(x),
\end{equation}
where $\ft(x)\in\C[x,x^{-1}]$ is not divisible by $x-1/x$ 
(i.e. $\ft(-1)\neq 0$ or $\ft(1)\neq 0$) and $2l+1\leq N_1+N_2$.

Let us denote $T=\min(N_1,N_2)$ and $T'=\max(N_1,N_2)>0$. 
Choosing $\epsilon =1$ if $T'=N_2$ and $\epsilon =-1$ otherwise we can 
rewrite $f_{N_1,N_2}(x)$ as
\begin{align*}
f_{N_1,N_2}(x)&=\frac{(x+\epsilon)^{2T'+1}(x-\epsilon)^{2T+1}}{x^{T'+T+1}}
    \nonumber\\
&=\left(x-\frac{1}{x}\right)^{2T+1}\left[x+2\epsilon+\frac{1}{x}\right]^{T'-T}.
\end{align*}
Below we consider separately the cases $l\leq T-1$ and $l\geq T$.\\

{\em Case 1: }$l\leq T-1.$ Let us write $\ft(x)$ in \eqref{5.9} as follows 
\begin{equation*}
\ft(x)=\sum_{j=0}^Sf_j\left(x+2\epsilon+\frac{1}{x}\right)^j.
\end{equation*}

{\em Case 1.a.}  Assume first that $f_0\neq 0$ and $T'>T$. Then, we can 
multiply $f(x)$ by 
$$\left(x-\frac{1}{x}\right)^{2(T-l)}
   \left(x+2\epsilon+\frac{1}{x}\right)^{T'-T-1}\in A$$
and we get
$$\sum_{j=0}^{S}f_j\left(x-\frac{1}{x}\right)^{2T+1}
\left(x+2\epsilon+\frac{1}{x}\right)^{j+T'-T-1}\in A.$$
Notice that for $j\geq 1$ the terms in the above sum are multiples of 
$f_{N_1,N_2}$ and therefore belong to $A$ by \eqref{5.5}. Thus, the term 
corresponding to $j=0$ also belongs to $A$ and since $f_{0}\neq 0$ we obtain 
$$\left(x-\frac{1}{x}\right)^{2T+1}
\left(x+2\epsilon+\frac{1}{x}\right)^{T'-T-1}=
\frac{(x+\epsilon)^{2T'-1}(x-\epsilon)^{2T+1}}{x^{T'+T}}\in A,$$
which is what we wanted to show.\\

{\em Case 1.b.\/}  Assume now that $f_0\neq 0$ but $T'=T$ 
(hence $\epsilon=1$), and let us multiply $f(x)$ by 
$$\left(x-\frac{1}{x}\right)^{2(T-l-1)}
   \left(x-2+\frac{1}{x}\right)\in A.$$
We have
$$\sum_{j=0}^{m}f_j\left(x-\frac{1}{x}\right)^{2T-1}
\left(x+2+\frac{1}{x}\right)^{j}\left(x-2+\frac{1}{x}\right)\in A.$$
Again all terms for $j\geq 1$ in the sum above are in $A$ and therefore, 
since $f_0\neq 0$, we get 
$$\left(x-\frac{1}{x}\right)^{2T-1}\left(x-2+\frac{1}{x}\right)=
\frac{(x+1)^{2T-1}(x-1)^{2T+1}}{x^{2T}}\in A.$$

{\em Case 1.c.\/} Let now $f_0=0$, i.e. $\ft(-\epsilon)=0$. This implies 
that $\ft(\epsilon)\neq 0$, because otherwise $(x-1/x)$ 
will divide $\ft(x)$ contrary to our factorization in \eqref{5.9}. Thus we can 
write $\ft(x)$ as
\begin{equation}\label{5.10}
\ft(x)=\sum_{j=0}^m\ft_j\left(x-2\epsilon+\frac{1}{x}\right)^j,
\end{equation}
with $\ft_0\neq 0$. 
Using now \eqref{5.10} and multiplying $f(x)$ by 
$$\left(x-2\epsilon+\frac{1}{x}\right)^{T-l-1}
\left(x+2\epsilon+\frac{1}{x}\right)^{T'-l}\in A$$
we get
\begin{equation*}
\sum_{j=0}^m\ft_j\left(x-\frac{1}{x}\right)^{2l+1}
    \left(x-2\epsilon+\frac{1}{x}\right)^{j+T-l-1}
    \left(x+2\epsilon+\frac{1}{x}\right)^{T'-l}\in A.
\end{equation*}
For $j\geq 1$, the $j$-th term in the above sum is
$$\left(x-2\epsilon+\frac{1}{x}\right)^{j-1}
\frac{(x-\epsilon)^{2T+1}(x+\epsilon)^{2T'+1}}{x^{T+T'+1}}\in A,$$
and therefore for $j=0$ we have 
$$\frac{(x-\epsilon)^{2T-1}(x+\epsilon)^{2T'+1}}{x^{T+T'}}\in A.$$

{\em Case 2.} Finally, if $l\geq T$ we can write $f(x)$ as 
\begin{equation}\label{5.11}
f(x)=\left(x-\frac{1}{x}\right)^{2T+1}\,\sum_{j=j_0}^{S}h_j
\left(x+2\epsilon+\frac{1}{x}\right)^j,
\end{equation}
with $h_{j_0}\neq 0$ and $S\leq T'-T-1$. Multiplying \eqref{5.11} by 
$\left(x+2\epsilon+\frac{1}{x}\right)^{T'-T-1-j_0}\in A$ 
we obtain as before that 
$$\frac{(x-\epsilon)^{2T+1}(x+\epsilon)^{2T'-1}}{x^{T+T'}}\in A,$$
which completes the proof of the lemma.
\end{proof}

\begin{Corollary}\label{co5.2}
If $A$ is a ring of Laurent polynomials in $x$, such that \\
$A_{N,N}\subset A$ for some $N\in\N_0$, then $A=A_{N_1,N_2}$ for 
some $N_1,N_2\in \N_0$.
\end{Corollary}

\begin{proof}
For every $f(x)\in A$ there exists a 
unique polynomial $p(x)$ such that $p(0)=0$ and $f(x)-p(x)\in\C[x+x^{-1}]$.
Since $x+x^{-1}\in A$ we see that $A$ is generated by $x+x^{-1}$ and the 
polynomial of minimal positive degree in $A\cap\C[x]$. The proof now follows 
immediately from \leref{le5.1}.
\end{proof}

Let us denote by $q_k(x)$ the odd polynomial of degree $2k+1$ given by
\begin{equation}\label{5.12}
q_k(x)=\sum_{j=0}^k\binom{2k+1}{j}(-1)^jx^{2k-2j+1}.
\end{equation}
Then
\begin{equation}\label{5.13}
f_{k,k}(x)=\left(x-\frac{1}{x}\right)^{2k+1}=q_k(x)-q_k(1/x).
\end{equation}
If we denote $u=x+x^{-1}$ then from \eqref{5.12} and \eqref{5.13} it is easy 
to see that
\begin{equation}\label{5.14}
q_{k+1}(x)=(u^2-4)q_k(x)+(-1)^{k+1}\binom{2k+1}{k}u.
\end{equation}
Using the last relation, one can deduce by induction on $k$ that 
\begin{equation}\label{5.15}
q_{k}(x)+q_{k}(1/x)=P_k(u),
\end{equation}
where
\begin{equation}\label{5.16}
\begin{split}
P_k(u)
&=\sum_{j=0}^k\frac{(-2)^j}{j!}\frac{(2k+1)!!}{(2k+1-2j)!!}\,u^{2k-2j+1}\\
&=\frac{(2k+1)!}{k!}
\sum_{j=0}^k\frac{(-1)^j}{j!}\frac{(k-j)!}{(2k+1-2j)!}\,u^{2k-2j+1},
\end{split}
\end{equation}
and $(2j+1)!!=1\cdot 3\cdots(2j+1)$.
\begin{Proposition}\label{pr5.3}
Let $L_{N_1,N_2}$ be the second-order difference operator constructed in 
\eqref{5.1} and $N=\max(N_1,N_2)$. Then for every $k\geq N$ we have
\begin{equation}\label{5.17}
P_k(L_{N_1,N_2})_+\in\cA_{N_1,N_2} \text{ and }
(L_{N_1,N_2}P_k(L_{N_1,N_2}))_+\in\cA_{N_1,N_2}.
\end{equation}
\end{Proposition}

\begin{proof} Clearly $P_k(L_{N_1,N_2})\in\cA_{N_1,N_2}$. From \eqref{5.2}, 
\eqref{5.5} and \eqref{5.13} it follows that $q_k(x)\in A_{N_1,N_2}$ for 
every $k\geq N$. This combined with equations \eqref{5.15} and \eqref{5.3a} 
shows that $P_k(L_{N_1,N_2})_+\in\cA_{N_1,N_2}$. Similar argument gives the 
second statement in \eqref{5.17}. Indeed, using that 
$$\left(x+\frac{1}{x}\right)f_{k,k}\in A_{N_1,N_2}\text{ for every }k\geq N,$$
and \eqref{5.13} one can deduce that $(x+1/x)q_k(x)\in A_{N_1,N_2}$ for all 
$k\geq N$, which combined with
$uP_k(u)=(x+1/x)q_k(x)+(-1)^k\binom{2k+1}{k}+O(1/x)$
gives the second part of \eqref{5.17}.
\end{proof}

\subsection{Characterization of $L_{N_1,N_2}$ in terms of the Toda flows}
\label{ss5.2}
In this subsection we prove that the operators 
$L_{N_1,N_2}$ can be characterized by the property \eqref{5.17}, or 
equivalently by the vanishing of an appropriate linear combination of the 
Toda flows after a particular point.
This is a discrete analog 
of the well-known fact that the rational solutions of the Korteweg-de Vries 
hierarchy are precisely the second-order differential operators, which are 
stationary under the KdV flows, after a particular point.
\begin{Theorem}\label{th5.4}
Let $L=E+b_n\Id+a_nE^{-1}$ be a nonconstant properly bordered second-order 
difference operator. Then the following conditions are equivalent.
\begin{itemize}
\item[(i)] The operator $L$ can be obtained by a sequence of Darboux 
transformations 
from the operator $L_{0}=E+E^{-1}$ at the end points $\pm 2$ of the spectrum, 
i.e. $L=L_{N_1,N_2}$ for some $N_1,N_2\in\N_0$ with $N_1+N_2>0$ and for 
specific values of the free parameters in the Darboux process \eqref{5.1}.
\item[(ii)] There exists $N\in\N$ such that 
\begin{equation}\label{5.18}
[P_k(L)_+,L]=[(LP_k(L))_+,L]=0 \text{ for every }k\geq N,
\end{equation}
where $P_k(u)$ is the polynomial defined by \eqref{5.16}.
\end{itemize}
\end{Theorem}

\begin{proof} The implication (i)$\Rightarrow$(ii) follows immediately from 
\prref{pr5.3}. Assume now that \eqref{5.18} holds for every $k\geq N$. Then 
$L$ belongs to a rank-one commutative ring of difference operators $\cA_L$ 
and $P_k(L)_+$, $(LP_k(L))_+\in\cA_L$ for $k\geq N$. Let $\Psi_n$ be the 
Baker function and $X=\Spec(\cA)\cup\{\Qip,\Qim\}$ be the complete curve. 
Then, there exists a function $\ff(P)$ on $X$ with simple poles at $\Qip$ and 
$\Qim$ such that
\begin{equation}\label{5.19}
L\Psi_n(P) =\ff(P)\Psi_n(P). 
\end{equation}
Let us pick local parameters $x^{-1}$ and  $y^{-1}$ near $\Qip$ and $\Qim$, 
respectively, such that 
\begin{equation}\label{5.20}
\ff(P)=\left\{
\text{\begin{tabular}{ll}
$x+1/x$ & when $P$ is in a neighborhood of $\Qip$\\
$y+1/y$ & when $P$ is in a neighborhood of $\Qim$.
\end{tabular}} 
\right.
\end{equation}
Then for each $k\geq N$ there exist functions $\ff_{2k+1}(P)$ and 
$\ff_{2k+2}(P)$ 
with poles only at $\Qip$ such that
\begin{subequations}\label{5.21}
\begin{align}
P_k(L)_+\Psi_n(P) &=\ff_{2k+1}(P)\Psi_n(P)\label{5.21a}\\
(LP_k(L))_+\Psi_n(P)& =\ff_{2k+2}(P)\Psi_n(P).\label{5.21b}
\end{align}
\end{subequations}
From \eqref{5.20} and \eqref{5.15} it follows that near $\Qip$ and $\Qim$ 
these functions have the following expansions:
\begin{subequations}\label{5.22}
\begin{align}
\ff^+_{2k+1}(x)&=q_k(x)+\sum_{j=1}^{\infty}\frac{\de_j^k}{x^j}\label{5.22a}\\
\ff^-_{2k+1}(y)&=\sum_{j=0}^{\infty}\frac{\dep_j^k}{y^j},\label{5.22b}
\end{align}
\end{subequations}
and 
\begin{subequations}\label{5.23}
\begin{align}
\ff^+_{2k+2}(x)&=\left(x+\frac{1}{x}\right)q_k(x)+(-1)^{k}\binom{2k+1}{k}
+\sum_{j=1}^{\infty}\frac{\ga_j^k}{x^j}\label{5.23a}\\
\ff^-_{2k+2}(y)&=\sum_{j=0}^{\infty}\frac{\gap_j^k}{y^j}.\label{5.23b}
\end{align}
\end{subequations}
Let us consider now the function 
$$\fg_{k}=\ff_{2k+1}\ff-\ff_{2k+2}\in A_L.$$
A straightforward computation using \eqref{5.20}, \eqref{5.22} and 
\eqref{5.23} shows that near $\Qip$ and $\Qim$ we have
\begin{subequations}\label{5.24}
\begin{align}
\fg^+_{k}(x)&=\de_1^k+(-1)^{k}\binom{2k+1}{k}+O(1/x)\label{5.24a}\\
\fg^-_{k}(y)&=\dep_0^ky+(\dep_1^k-\gap_0^k)+O(1/y).\label{5.24b}
\end{align}
\end{subequations}
If $\dep_0^k\neq 0$, the function $\fg_k$ will correspond to a difference 
operator from $\cA_L$ with support $[-1,0]$, which would imply that $L$ is a 
constant coefficient operator, contrary to our assumption. Thus 
\begin{equation}\label{5.25}
\dep_0^k=0
\end{equation}
and $\fg_k(P)$ is a constant (depending on $k$).
Next, we define 
$$\fhh_k=\ff_{2k+3}-(\ff^2-4)\ff_{2k+1}+(-1)^k\binom{2k+1}{k}\ff\in A_L.$$
We have 
\begin{subequations}\label{5.26}
\begin{align}
\fhh^+_{k}(x)&=-\de_1^kx-\de_2^k+O(1/x)\label{5.26a}\\
\fhh^-_{k}(y)&=\left(-\dep_1^k+(-1)^k\binom{2k+1}{k}\right)y
-\dep_2^k+O(1/y),\label{5.26b}
\end{align}
\end{subequations}
and the same argument as above shows that $\fhh_k+\de_1^k\ff+\de_2^k=0$, i.e.
\begin{equation}\label{5.27}
\ff_{2k+3}-(\ff^2-4)\ff_{2k+1}+(-1)^k\binom{2k+1}{k}\ff+\de_1^k\ff+\de_2^k=0.
\end{equation}
In particular, this equality implies that
\begin{align}
&\de_1^k=\dep_1^k+(-1)^{k+1}\binom{2k+1}{k}\label{5.28}\\
&\de_2^k=\dep_2^k.\label{5.29}
\end{align}
We want to show now that for every $k\geq N$, around $\Qim$ we have
\begin{equation}\label{5.30}
\ff^-_{2k+1}(y)=q_k(1/y)+\sum_{j=1}^{\infty}\frac{\de_j^k}{y^j}.
\end{equation}
From equations \eqref{5.14}, \eqref{5.20}, \eqref{5.22a} and \eqref{5.27} we 
deduce that coefficients $\de^k_j$ in the expansion of the function 
$\ff^+_{2k+1}(x)-q_k(x)=\sum_{j=1}^{\infty}\de_j^k/x^j$ satisfy the 
following recurrence relations
\begin{subequations}\label{5.31}
\begin{align}
&\de_3^k=\de_1^{k+1}+3\de_1^k \label{5.31a}\\
&\de_{j+2}^{k}=\de_j^{k+1}+2\de_j^k-\de_{j-2}^k \text{ for }j\geq 2.
                                        \label{5.31b}
\end{align}
\end{subequations}
The same argument shows that the same relations will be satisfied by 
the coefficients in the expansion of the function $\ff^-_{2k+1}(y)-q_k(1/y)$. 
Thus, to prove that \eqref{5.30} holds, it is enough to show that the 
coefficient of $1/x^j$ in $\ff^+_{2k+1}(x)-q_k(x)$ is equal to the 
coefficient of $1/y^j$ in $\ff^-_{2k+1}(y)-q_k(1/y)$ for $j=0,1,2$ and every 
$k\geq N$. It is easy to see that the equality of these three coefficients 
is equivalent to equations \eqref{5.25}, \eqref{5.28} and \eqref{5.29}, 
completing the proof of \eqref{5.30}.

Next, we denote 
$$\fp_{k}=2\ff_{2k+1}-P_k(\ff)\in A_L.$$
Using equations \eqref{5.13}, \eqref{5.15}, \eqref{5.22a} and \eqref{5.30} 
we see that $\fp_k(P)$ has the following expansions near $\Qip$ and $\Qim$
\begin{subequations}\label{5.32}
\begin{align}
\fp^+_{k}(x)&=q_k(x)-q_k(1/x)
+2\sum_{j=1}^{\infty}\frac{\de_j^k}{x^j}\nonumber\\
&=\left(x-\frac{1}{x}\right)^{2k+1}
+2\sum_{j=1}^{\infty}\frac{\de_j^k}{x^j}\label{5.32a}\\
\fp^-_{k}(y)&=-q_k(y)+q_k(1/y)
+2\sum_{j=1}^{\infty}\frac{\de_j^k}{y^j}\nonumber\\
&=-\left(y-\frac{1}{y}\right)^{2k+1}
+2\sum_{j=1}^{\infty}\frac{\de_j^k}{y^j}.\label{5.32b}
\end{align}
\end{subequations}
We now use the fact that $\fp_k$ and $\ff$ satisfy an algebraic relation of 
the form 
\begin{equation}\label{5.33}
\fp_{k}^2=\fp_kT_1(\ff)+T_2(\ff),
\end{equation}
for some polynomials $T_1$ and $T_2$. It is easy to show that 
$T_1(\ff)=0$. Indeed, if we assume that $T_1(\ff)=r\ff^s+\cdots$ for some 
nonzero constant $r$, then the function $\fp_kT_1(\ff)$ has the following 
expansions near $\Qip$ and $\Qim$
\begin{equation*}
\fp_kT_1(\ff)=\left\{
\text{\begin{tabular}{ll}
$rx^{2k+1+s}+O(x^{2k+s})$ & in a neighborhood of $\Qip$\\
$-ry^{2k+1+s}+O(y^{2k+s})$ & in a neighborhood of $\Qim$.
\end{tabular}} 
\right.
\end{equation*}
On the other hand, from equations \eqref{5.32} it is clear that if 
\begin{equation*}
\fp_{k}^2-T_2(\ff)=\left\{
\text{\begin{tabular}{ll}
$rx^{l}+O(x^{l-1})$ & in a neighborhood of $\Qip$\\
$-ry^{l}+O(y^{l-1})$ & in a neighborhood of $\Qim$,
\end{tabular}} 
\right.
\end{equation*}
for some nonzero constant $r$, then $l\leq 2k$, leading to a contradiction. 
Thus $T_1(\ff)=0$ and therefore \eqref{5.33} reduces to
\begin{equation}\label{5.34}
\fp_{k}^2=T_2(\ff).
\end{equation}
We can show now that $\de_j^k=0$ for all $j\in\N$ which implies that
$T_2(\ff)=(\ff^2-4)^{2k+1}$. Indeed, let $\de_{j_0}^k\neq 0$ for some 
$j_0\in\N$ and let $j_0$ be the minimal possible. Then we can rewrite 
\eqref{5.34} as 
\begin{equation}\label{5.35}
\fp_{k}^2-(\ff^2-4)^{2k+1}=T_2(\ff)-(\ff^2-4)^{2k+1}.
\end{equation}
The left-hand side of \eqref{5.35} has the following expansions near 
$\Qip$ and $\Qim$
\begin{equation*}
\fp_{k}^2-(\ff^2-4)^{2k+1}=\left\{
\text{\begin{tabular}{ll}
$4\de_{j_0}^kx^{2k+1-j_0}+O(x^{2k-j_0})$ & in a neighborhood of $\Qip$\\
$-4\de_{j_0}^ky^{2k+1-j_0}+O(y^{2k-j_0})$& in a neighborhood of $\Qim$.
\end{tabular}} 
\right.
\end{equation*}
Again we get a contradiction because clearly the highest coefficients in the 
expansion of 
$T_2(\ff)-(\ff^2-4)^{2k+1}$ near $\Qip$ and $\Qim$ must be equal.
Thus \eqref{5.34} becomes
\begin{equation}\label{5.36}
\fp_{k}^2=(\ff^2-4)^{2k+1}.
\end{equation}
Notice that the curve given by the last equation is exactly the spectral curve 
$\Spec(\cA_{k,k})$ in \eqref{5.4}. Thus, we can conclude that the spectral 
curve $X$ is rational and choosing an appropriate parametrization we have
$$A_{N,N}\subset A_L\subset \C[x,x^{-1}].$$
From \coref{co5.2} it follows that $A=A_{N_1,N_2}$ for some $N_1,N_2\in \N_0$ 
with $N_1+N_2>0$ (because $L$ is nonconstant). Moreover, as we saw at the 
beginning of the section, equation \eqref{5.7} must hold. Let $\Psi_n(x)$ be 
the Baker function for $\cA_{N_1,N_2}$, i.e. the function in formulas 
\eqref{5.3a}-\eqref{5.3b}. Then, the Baker function for $\cA_L$ is 
$\tilde{\Psi}_n(x)=c^{-n}\Psi_n(x)$. Thus 
$$L\tilde{\Psi}_n(x)
=\left(\frac{1}{c}\left(x+\frac{1}{x}\right)+d\right)\tilde{\Psi}_n(x).$$
Equation \eqref{5.36} shows that for $k$ large enough the function 
$$\left[\left(\frac{1}{c}\left(x+\frac{1}{x}\right)+d\right)^2
-4\right]^{2k+1}$$
must be the square of a rational function. It is easy to see that this can 
only happen when $d=0$ and $c=\pm 1$, which means that $L=L_{N_1,N_2}$ or 
$L=L_{N_2,N_1}$ finishing the proof.
\end{proof}

\begin{Remark} \label{re5.5}
We can easily reformulate \thref{th5.4} in terms of the vector 
fields $\X_k$ of the Toda lattice hierarchy \eqref{2.1}. For every $k\in\N$ 
let
\begin{equation}\label{5.37}
\varepsilon_k=
\begin{cases} 0 & \text{if $k$ is odd}\\ 
1 & \text{if $k$ is even,}
\end{cases}
\end{equation}
and let us define
\begin{equation}\label{5.38}
\Y_k(L)=\frac{(k-\varepsilon_k)!}{(\lfloor(k-1)/2\rfloor)!}
\sum_{j=0}^{\lfloor(k-1)/2\rfloor}\frac{(-1)^j}{j!}
\frac{(\lfloor(k-1)/2\rfloor-j)!}{(k-\varepsilon_k-2j)!}\,\X_{k-2j}(L),
\end{equation}
where $\X_j(L)=[(L^j)_+,L]$.
Then the operators $L_{N_1,N_2}$ with $N_1+N_2>0$ can be characterized as the 
only nonconstant properly bordered second-order difference operators of the 
form $L=E+b_n\Id+a_nE^{-1}$ satisfying the constraints
\begin{equation}\label{5.39}
\Y_k(L)=0 \text{ for every $k$ large enough.}
\end{equation}
\end{Remark}

We end this section by giving an explicit formula for $\Y_k$ in terms of the 
vector fields $\X'_j$ defined in \thref{th4.4}. The proposition below will 
be needed in the next section when we want to characterize the operators 
having specific heat kernel expansions.

\begin{Proposition}\label{pr5.6} 
The vector fields $\Y_k$ defined by \eqref{5.38} can be rewritten as a 
linear combination of the vector fields $\X'_{k}$, corresponding to the flows 
of the system \eqref{4.3}, as follows
\begin{equation}\label{5.40}
\Y_k=\sum_{l=0}^{\lfloor(k-1)/2\rfloor}(-1)^l
\left(1-\frac{2l}{k}\varepsilon_k\right)\binom{k}{l}\X'_{k-2l}.
\end{equation}
\end{Proposition}

\begin{proof}
Using \eqref{4.7} and \eqref{5.38} we get
 \begin{align*}
\Y_k&=\frac{(k-\varepsilon_k)!}{(\lfloor(k-1)/2\rfloor)!}
\sum_{j=0}^{\lfloor(k-1)/2\rfloor}\frac{(-1)^j}{j!}
\frac{(\lfloor(k-1)/2\rfloor-j)!}{(k-\varepsilon_k-2j)!}\\
&\qquad\qquad \times
\sum_{i=0}^{\lfloor(k-1)/2\rfloor-j}\binom{k-2j}{i}\X'_{k-2i-2j}\\
&=\sum_{l=0}^{\lfloor(k-1)/2\rfloor}R_{k,l}\X'_{k-2l},
\end{align*}
where
\begin{equation}\label{5.41}
R_{k,l}=\frac{(k-\varepsilon_k)!}{(\lfloor(k-1)/2\rfloor)!}
\sum_{j=0}^{l}\frac{(-1)^j}{j!}
\frac{(\lfloor(k-1)/2\rfloor-j)!(k-2j)!}{(k-\varepsilon_k-2j)!(l-j)!(k-l-j)!}.
\end{equation}
We consider now separately the cases when $k$ is odd or even. 

{\it Case 1.} Assume first that $k$ is odd, i.e. $k=2s+1$. Then 
$\varepsilon_k=0$ and \eqref{5.41} gives
\begin{align*}
R_{2s+1,l}&=\frac{(2s+1)!}{s!}\sum_{j=0}^{l}\frac{(-1)^j}{j!}
\frac{(s-j)!}{(l-j)!(2s+1-l-j)!}\\
&=\binom{2s+1}{l}{}_2F_1\Big(
\begin{matrix} -l,-2s-1+l\\
-s\end{matrix}\; ;1\Big)\\
&=\binom{2s+1}{l}\frac{(s+1-l)_l}{(-s)_l}=(-1)^l\binom{2s+1}{l},
\end{align*}
where in the last line we used the Chu-Vandermonde formula to evaluate 
the ${}_2F_1$ and $(a)_l=a(a+1)\cdots(a+l-1)$ denotes the shifted factorial.
This completes the proof in the case when $k$ is odd.

{\it Case 2.} Assume now that $k$ is even, i.e. $k=2s+2$ and therefore
$\varepsilon_k=1$. A similar computation as above gives
\begin{align*}
R_{2s+2,l}&=2\frac{(2s+1)!}{s!}\sum_{j=0}^{l}\frac{(-1)^j}{j!}
\frac{(s+1-j)!}{(l-j)!(2s+2-l-j)!}\\
&=\binom{2s+2}{l}{}_2F_1\Big(
\begin{matrix} -l,-2s-2+l\\
-s-1\end{matrix}\; ;1\Big)\\
&=\binom{2s+2}{l}\frac{(s+1-l)_l}{(-s-1)_l}=(-1)^l\,
\frac{s-l+1}{s+1}\binom{2s+2}{l},
\end{align*}
which completes the proof.
\end{proof}

\section{Finite heat kernel expansions}\label{se6}
In this section we prove that the operators $L_{N_1,N_2}$ constructed in 
the previous section can be characterized as the only operators for which the 
heat kernel can be written as a sum of only two Bessel functions with 
polynomial (in $t$) coefficients.

We work below with series of the form 
\begin{equation}\label{6.1}
\sum_{r\leq k}p(r)I_r(2t),
\end{equation}
where $p(r)$ is a polynomial in $r$, and before we state the main result, 
we establish several properties of the series \eqref{6.1}. Using \eqref{1.7} 
we see that for $r\geq 0$ we have
\begin{equation*}
|I_{-r}(2t)|\leq 
\frac{|t|^r}{r!}\sum_{j=0}^{\infty}\frac{|t|^{2j}}{j!}
\leq \frac{|t|^r}{r!}e^{|t|^2}.
\end{equation*}
Thus if $|t|\leq T$, then $|I_{-r}(2t)|\leq \frac{|T|^r}{r!}e^{|T|^2}$. Since 
for every polynomial $p(r)$ and for every positive constant $T$ the series 
$\sum_{r=0}^{\infty}\frac{|p(-r)|}{r!}T^r$ converges we conclude that the 
series \eqref{6.1} converges absolutely and uniformly on bounded sets.

The next lemma allows us to identify polynomials $p(r)$ for which the 
sum of the series \eqref{6.1} can be written in a simple closed form.

\begin{Lemma} \label{le6.1}
For every $m\in\N$ and  $k\in \Z$ we have
\begin{equation}\label{6.2}
t^{m}I_k(2t)=(-1)^m
\sum_{\begin{subarray}{c}i\leq k\\ i\equiv k+m\; ({\mathrm{mod}}\;2) 
\end{subarray}}A^{m,k}_i\,I_{i}(2t),
\end{equation}
where
\begin{equation}\label{6.3}
A^{m,k}_i=\frac{i}{4^{m-1}(m-1)!}
\prod_{\begin{subarray}{c}|j|<m\\ j\equiv m\; ({\mathrm{mod}}\;2) 
\end{subarray}}((k+j)^2-i^2).
\end{equation}
\end{Lemma}
\begin{proof} The proof can be obtained by induction on $m$ using 
the identity
\begin{equation}\label{6.4}
t(I_{k-1}(2t)-I_{k+1}(2t))=kI_{k}(2t).
\end{equation}
\end{proof}
Notice that $A^{m,k}_i$ is an odd polynomial in $i$ of degree $2m-1$. Thus, 
as an immediate application of the above lemma, we obtain the following 
corollary.

\begin{Corollary} \label{co6.2}
Let $p(i)$ be a function which is an odd polynomial in $i$, when $i$ is even 
or odd.\footnote{Equivalently, we can say that 
$p(i)=\tilde{p}(i)+(-1)^i\hat{p}(i)$, where $\tilde{p}(i)$ and $\hat{p}(i)$ 
are odd polynomials in $i$.} Then 
$$\sum_{i<k}p(i)I_i(2t)=p_1(t)I_{k}(2t)+p_2(t)I_{k-1}(2t),$$
where $p_1(t)$ and $p_2(t)$ are polynomials of $t$ such that 
$p_1(0)=p_2(0)=0$.
\end{Corollary}

We can now formulate and prove the main result in this section.

\begin{Theorem} \label{th6.3}
Let $L=E+b_n\Id+a_nE^{-1}$ be a properly bordered second-order difference 
operator. Then the following conditions are equivalent.
\begin{itemize}
\item[(i)] The operator $L$ can be obtained by a sequence of Darboux 
transformations from the operator $L_{0}=E+E^{-1}$ at the end points $\pm 2$
of the spectrum, 
i.e. $L=L_{N_1,N_2}$ for some $N_1,N_2\in\N_0$ and for some specific values 
of the free parameters in the Darboux process \eqref{5.1}.
\item[(ii)] The fundamental solution of the discrete heat equation 
\eqref{1.2} can be written as
\begin{equation}\label{6.5}
u(n,m;t)=(1+p_1(n,m;t))I_{n-m}(2t)+p_2(n,m;t)I_{n-m-1}(2t),
\end{equation}
where $p_1(n,m;t)$ and $p_2(n,m;t)$ are polynomials in $t$ with coefficients 
depending on $n$ and $m$, such that $p_1(n,m;0)=p_2(n,m;0)=0$.
\end{itemize}
\end{Theorem}

\subsection{Finiteness of the heat kernel for the operators $L_{N_1,N_2}$}
\label{ss6.1}
The implication (i)$\Rightarrow$(ii) is 
essentially proved in \cite{GI}, except the fact that exactly two Bessel 
functions ($I_{n-m}$ and $I_{n-m-1}$) are enough. 
This can be deduced from the arguments given there combined with 
\coref{co6.2}. We explain below the main steps of the proof together with 
the essential ingredients from \cite{GI,HI2}.

We consider the operator $L_{N_1,N_2}$ obtained by the sequence of 
Darboux transformations \eqref{5.1} and the corresponding 
maximal commutative ring $\cA_{N_1,N_2}$ of difference operators that 
contains $L_{N_1,N_2}$. From \eqref{5.1} one can deduce that 
$L_{N_1,N_2}$ and $L_0=E+E^{-1}$ are related by the intertwining 
relation
\begin{equation}\label{6.6}
Q L_0 =L_{N_1,N_2}Q, 
\end{equation}
where
\begin{equation*}
Q=\cQ_{N_1+N_2-1}\cdots \cQ_1\cQ_0.
\end{equation*}
Equation \eqref{6.6} implies that $\ker Q$ is preserved by $L_0$, i.e. 
$L_0(\ker Q)\subset \ker Q$. Conversely, one can show by induction that if 
two operators $L_0$ and $L=L_{N_1,N_2}$ are related by \eqref{6.6} for some
difference operator $Q$, then $L$ can be obtained by a sequence of Darboux 
transformations from $L_0$. The fact that we iterate the Darboux 
transformation only at $\pm 2$ means that the operator $L_0$ restricted to 
$\ker Q$ has $2$ and $-2$ as eigenvalues with multiplicities $N_1$ and $N_2$, 
respectively. This allows us to reconstruct $Q$ explicitly from its kernel 
as follows. We define functions $\phi^+_1,\phi^+_2,\dots,\phi^+_{N_1}$ and 
$\phi^-_1,\phi^-_2,\dots,\phi^-_{N_2}$ such that
\begin{align*}
&(L_0-2\Id)\phi^+_j=\phi^+_{j-1}, \text{ for }j=1,2,\dots,N_1\\
&(L_0+2\Id)\phi^-_j=\phi^-_{j-1}, \text{ for }j=1,2,\dots,N_2,
\end{align*}
with the convention that $\phi^+_{0}=\phi^-_{0}=0$. Let $\Wr$ denote 
the discrete Wronskian (Casorati determinant) with respect to the variable $n$
$$\Wr(f_1,f_2,\dots,f_k)=\det(\Delta^{i-1}f_j)_{1\leq i,j\leq k}.$$
Then the operator $Q$, normalized to be monic, is defined by 
\begin{equation}\label{6.7}
Q(f)=
\frac{\Wr(\phi^+_1,\dots,\phi^+_{N_1},\phi^-_1,\dots,\phi^-_{N_2},f)}
{\Wr(\phi^+_1,\dots,\phi^+_{N_1},\phi^-_1,\dots,\phi^-_{N_2})}.
\end{equation}
The Baker function for the ring $\cA_{N_1,N_2}$ can be written in terms of 
the operator $Q$ by
\begin{equation}\label{6.8}
\Psi_n(x)=\frac{1}{(x-1)^{N_1}(x+1)^{N_2}}Q(x^n). 
\end{equation}
Using the explicit form of the functions $\phi^{\pm}_j$, equations 
\eqref{6.7} and \eqref{6.8} one can show that $\Psi_n(1/x)$ and 
$x\Psi^*_{n+1}(x)$ are equal up to a multiplicative constant independent of 
$x$, i.e. 
\begin{equation}\label{6.9}
\Psi_n(1/x)=c_nx\Psi_{n+1}^*(x). 
\end{equation}
Finally, we can use the above information to prove that 
$\Psi_n(x)$ and $\Psi_n^*(x)$ satisfy 
the orthogonality relation
\begin{equation}\label{6.10}
\frac{1}{2\pi i}\oint_C \Psi_n(x)\Psi_{m+1}^*(x)\,dx=\delta_{n,m},
\end{equation}
where  $C$ is a simple closed contour around the origin, avoiding the points 
$x=\pm 1$. The proof of \eqref{6.10} can be obtained as follows. Notice first 
that $\Psi_n(x)$ and $\Psi_{m+1}^*(x)$ have poles only at $\pm 1$. 
However the spectral curve has cusps at these points and since the 
differential $\Psi_n(x)\Psi_{m+1}^*(x)\,dx$ is regular on the affine curve 
$\Spec(\cA_{N_1,N_2})$ we deduce that residues at $x=\pm1$ are equal to zero. 
Using the explicit formulas for $\Psi_n(x)$ and $\Psi_{m+1}^*(x)$ we see that 
expanding  around $x=0$ for $m\leq n$ we have
$$\Psi_n(x)\Psi_{m+1}^*(x)=\frac{1}{x}\left(\delta_{n,m}+O(x)\right),$$
which establishes  \eqref{6.10} for  $m\leq n$. When $m>n$ we replace $x$ by 
$1/x$ in \eqref{6.10} and applying \eqref{6.9} we obtain zero by the bilinear 
identity \eqref{2.9}. We refer the reader to \cite{HI2} for detailed proofs of 
all statements in the above construction. 

From \eqref{5.3a} and \eqref{6.10} it follows that the fundamental solution 
for $L_{N_1,N_2}$ can be written as
\begin{equation}\label{6.11}
u(n,m;t)=\frac{1}{2\pi i}\oint_C e^{t(x+\frac{1}{x})}
\Psi_n(x)\Psi_{m+1}^*(x)\,dx.
\end{equation}
Since 
$$e^{t(x+\frac{1}{x})}=\sum_{k\in\Z}x^kI_k(2t)$$
we can deduce from \eqref{6.11} that 
\begin{equation*}
u(n,m;t)=\sum_{k=0}^{\infty}\alpha_k(n,m)I_{n-m-k}(2t),
\end{equation*}
where
\begin{equation}\label{6.12}
\alpha_k(n,m)=\frac{1}{2\pi i}\oint_C x^{k-n+m}\Psi_n(x)\Psi_{m+1}^*(x)\,dx.
\end{equation}
Using the last formula, we prove that for $k\geq 1$, running over the even 
or the odd integers $\alpha_k(n,m)$ is an odd function of $n-m-k$ with 
coefficients depending on $n$ and $m$. The statement then will follow from 
\coref{co6.2}.

The idea of the proof is to write $x^{k-n+m}$ in \eqref{6.12} as 
$[x^{k-n+m}-p(x,n-m-k)]+p(x,n-m-k)$ for appropriate $p(x,n-m-k)$ which is 
a Laurent polynomial of $x$, and an odd polynomial of $n-m-k$.
We want to pick $p(x,n-m-k)$ so that 
$f(x)=x^{k-n+m}-p(x,n-m-k)\in A_{N_1,N_2}$. Then there exists a difference 
operator $L_f=\sum_{l=K_-}^{K_+}\mu_{l}(n)E^l\in\cA_{N_1,N_2}$ and therefore 
$$f(x)\Psi_n(x)=L_f\Psi_n(x)=\sum_{l=K_-}^{K_+}\mu_l(n)\Psi_{n+l}(x).$$
If the interval $[n+K_-,n+K_+]$ does not contain $m$  
we deduce from \eqref{6.10} that
\begin{equation*}
\frac{1}{2\pi i}\oint_C f(x)\Psi_n(x)\Psi_{m+1}^*(x)\,dx=0,
\end{equation*}
and therefore formula \eqref{6.12} will give 
\begin{equation*}
\alpha_k(n,m)=\frac{1}{2\pi i}\oint_C p(x,n-m-k) \Psi_n(x)\Psi_{m+1}^*(x)\,dx,
\end{equation*}
completing the proof. The main problem now is to construct a class of 
Laurent polynomials in $A_{N_1,N_2}$, which allows to implement 
the above idea. The key ingredient is the following proposition established 
in \cite{GI}.

\begin{Proposition}\label{pr6.4}
Let $N\geq\max(N_1,N_2)$, and let $l_0,l_1,\dots,l_N$ be 
distinct nonzero integers, such that $l_j\equiv l_k\pmod 2$ and 
$l_j+l_k\neq 0$, for $0\leq j,k\leq N$. Then
\begin{equation}\label{6.13}
\sum_{k=0}^N\frac{x^{l_k}}{l_k\prod_{j\neq k}(l_k^2-l_j^2)}
\in A_{N_1,N_2}.
\end{equation}
\end{Proposition}

To complete the proof, we fix $n$ and $m$, and 
we choose $\epsilon =1$ or $\epsilon =2$ so that $k\equiv\epsilon\mod 2$. 
Let $N=\max(N_1,N_2)$ and let us denote 
\begin{equation*}
h^s_{J}(j)=\frac{j}{J+2s}
\prod_{\begin{subarray}{c} l=0\\ l\neq s\end{subarray}}^{N-1}
\frac{j^2-(J+2l)^2}{(J+2s)^2-(J+2l)^2}.
\end{equation*}
Clearly, $h^s_{J}(j)$ is an odd polynomial of $j$. Now we can rewrite 
\eqref{6.12} as follows
\begin{align}
&\alpha_k(n,m)=\frac{1}{2\pi i}\oint_C 
\left(x^{k-n+m}-\sum_{s=0}^{N-1}h^s_{J}(k-n+m)x^{J+2s}\right)
           \Psi_n(x)\Psi_{m+1}^*(x)\,dx \nonumber\\
&\quad +\sum_{s=0}^{N-1}\frac{h^s_{J}(k-n+m)}{2\pi i}
\oint_C x^{J+2s}\Psi_n(x)\Psi_{m+1}^*(x)\,dx.
\label{6.14}
\end{align}
We need to define $J$ depending only on $n$ and $m$ so that 
\begin{equation}\label{6.15}
\frac{1}{2\pi i}\oint_C 
\left(x^{k-n+m}-\sum_{s=0}^{N-1}h^s_{J}(k-n+m)x^{J+2s}\right)
           \Psi_n(x)\Psi_{m+1}^*(x)\,dx=0,
\end{equation}
for every $k$, because then the right-hand side of equation \eqref{6.14} will 
clearly be an odd polynomial of $n-m-k$ with 
coefficients depending on $n$ and $m$.

Assume first that $n\leq m$ and take $J=m-n+\epsilon$. 
If $1 \leq k\leq 2N$, then $k-n+m=J+2s$ for some $s\in\{0,1,\dots,N-1\}$ 
and \eqref{6.15} is obvious because 
the polynomial 
$$f(x)=x^{k-n+m}-\sum_{s=0}^{N-1}h^s_{J}(k-n+m)x^{J+2s}$$ 
is identically equal to $0$. If $k>2N$, then $f(x)\in A_{N_1,N_2}$ 
by \prref{pr6.4} and therefore $f(x)\Psi_n(x)$ can be written 
as a linear combination of 
$\{\Psi_{k+m}(x),\Psi_{k+m-1}(x),\dots,\Psi_{m+\epsilon}(x)\}$ 
and thus \eqref{6.15} follows from \eqref{6.10}.

If $n>m$ we can use a similar argument by taking $J=n-m+\epsilon$. This 
completes the proof of the implication (i)$\Rightarrow$(ii). 
\qed

\subsection{Characterization of $L_{N_1,N_2}$ in terms of the 
heat kernel}\label{ss6.2}
In this subsection we prove that (ii) implies (i) in \thref{th6.3}. 
Assume first that $L$ is a nonconstant second-order difference operator. The 
strategy of the proof is to show that \eqref{5.39} holds using \eqref{5.40} 
and then apply \thref{th5.4} and \reref{re5.5}.

From \eqref{6.5} we see that
\begin{equation}\label{6.16}
u(n,n;t)=I_0(2t)+p_1(n,n;t)I_{0}(2t)+p_2(n,n;t)I_{-1}(2t).
\end{equation}
On the other hand, \eqref{1.4} gives
\begin{equation}\label{6.17}
u(n,n;t)=I_0(2t)+\sum_{k=1}^{\infty}\alpha_k(n,n)I_{-k}(2t).
\end{equation}
Using equations \eqref{6.16}-\eqref{6.17}, the fact that the coefficients 
$\alpha_k(n,n)$ in 
the expansion \eqref{6.17} are uniquely determined by $u(n,n;t)$ and 
\leref{le6.1} we see that for $k\geq 1$, running over the 
even or the odd integers, $\alpha_{k}(n,n)$ is an odd polynomial in $k$ with 
coefficients depending on $n$. Similarly, using again \eqref{6.5} and 
writing $u(n+1,n;t)$ as
\begin{equation*}
u(n+1,n;t)=I_1(2t)+p_2(n+1,n;t)I_{0}(2t)+p_1(n+1,n;t)I_{-1}(2t),
\end{equation*}
we can conclude that for $k\geq 1$, running over the 
even or the odd integers, $\alpha_{k+1}(n+1,n)$ is an odd polynomial in $k$ 
with coefficients depending on $n$. 

Let $2N-1$ be the maximal degree of the four polynomials of $k$: 
$\alpha_{k}(n,n)$ when $k$ is odd/even and $\alpha_{k+1}(n+1,n)$ when $k$ is 
odd/even. We show below that $\Y_k(L)=0$ for all $k\geq 2N+1$. From 
\eqref{5.40} and \eqref{4.3} it follows that it suffices to prove that if 
$f(x)$ is an odd polynomial of degree at most $2N-1$, then
\begin{equation}\label{6.18}
\sum_{l=0}^{\lfloor(k-1)/2\rfloor}(-1)^l
\left(1-\frac{2l}{k}\varepsilon_k\right)\binom{k}{l}f(k-2l)=0,
\end{equation}
for every $k\geq 2N+1$, where $\varepsilon_k$ is defined by \eqref{5.37}.\\

If $k$ is odd, i.e. $k=2s+1$, equation \eqref{6.18} reduces to 
\begin{equation}\label{6.19}
\sum_{l=0}^{s}(-1)^l\binom{2s+1}{l}f(2s+1-2l)=0.
\end{equation}
Since $f$ is odd, we see that the left-hand side of \eqref{6.19} is
equal to
\begin{equation*}
\frac{1}{2}\sum_{l=0}^{2s+1}(-1)^l\binom{2s+1}{l}f(2s+1-2l)
\end{equation*}
and therefore \eqref{6.19} is equivalent to
\begin{equation}\label{6.20}
\sum_{l=0}^{2s+1}(-1)^l\binom{2s+1}{l}f(2s+1-2l)=0.
\end{equation}
Equation \eqref{6.20} will follow immediately if we can show that for every 
polynomial $F(x)$ of degree less that $j$, we have
\begin{equation}\label{6.21}
\sum_{l=0}^{j}(-1)^l\binom{j}{l}F(l)=0.
\end{equation}
The proof of the last identity is straightforward:
\begin{align*}
&\sum_{l=0}^{j}(-1)^l\binom{j}{l}F(l)
=\left[F(\pd_z)\sum_{l=0}^{j}(-1)^l\binom{j}{l}e^{lz}\right]\Bigg\vert_{z=0}\\
&\qquad =\left[F(\pd_z)(1-e^z)^j\right]\big\vert_{z=0}=0.
\end{align*}
If $k$ is even, i.e. $k=2s+2$, then \eqref{6.18} is equivalent to
\begin{align*}
&\sum_{l=0}^{s}(-1)^l\binom{2s+2}{l}(2s+2-2l)f(2s+2-2l) \\
&\qquad =\frac{1}{2}
\sum_{l=0}^{2s+2}(-1)^l\binom{2s+2}{l}(2s+2-2l)f(2s+2-2l)=0,
\end{align*}
which follows again from \eqref{6.21}.

It remains to prove the statement when $L$ has constant coefficients, 
i.e. $L=E+b\Id+aE^{-1}$, where $a$ and $b$ are constants and $a\neq 0$. 
In this case, we can write an explicit formula for $u(n,m;t)$ and 
calculate $\alpha_k(n,m)$. For the fundamental solution we obtain 
\begin{equation}\label{6.22}
u(n,m;t)=\frac{1}{2\pi i}\oint_C e^{\left(z+b+\frac{a}{z}\right)t}z^{n-m-1}dz,
\end{equation}
where $C$ is a simple closed contour around the origin. Let $w$ be such that 
$z+b+\frac{a}{z}=w+\frac{1}{w}$ and $w\rightarrow 0$ as $z\rightarrow 0$. 
Using \eqref{6.22}, a short computation shows that 
\begin{equation}\label{6.23}
\alpha_k(0,0)=\res_{w=0}
\left(\frac{(1-w^2)w^{-1-k}}{\sqrt{\fq(w)}}\right),
\end{equation}
where
$$\fq(w)=1-2bw+(2+b^2-4a)w^2-2bw^3+w^4.$$
In other words, $\alpha_k(0,0)$ are the coefficients in the expansion 
of the function $(1-w^2)/\sqrt{\fq(w)}$ around $w=0$. We know from 
\leref{le6.1} that if \eqref{6.5} holds for $m=n=0$ then $\alpha_k(0,0)$ 
must be an odd polynomial in $k$ for $k$ odd or even. Let 
\begin{equation*}
\alpha_k(0,0)=
\begin{cases} \beta_1(k) & \text{when $k$ is odd}\\ 
\beta_2(k) & \text{when $k$ is even}
\end{cases}
\end{equation*}
Then
\begin{align*}
&\sum_{k=1}^{\infty}\alpha_k(0,0)w^{k}
=\beta_1(w\pd_w)\sum_{j=1}^{\infty}w^{2j-1}
+\beta_2(w\pd_w)\sum_{j=1}^{\infty}w^{2j}\\
&\quad =\beta_1(w\pd_w) \frac{w}{1-w^2} 
+ \beta_2(w\pd_w) \frac{w^2}{1-w^2} 
=\frac{\text{polynomial of $w$}}{(1-w^2)^K},
\end{align*}
showing that $\sum_{k=1}^{\infty}\alpha_k(0,0)w^{k}$ 
must be a rational function of $w$ with denominator 
having zeros only at $w=\pm 1$. This implies that the only possible 
choices for $\fq(w)$ in \eqref{6.23} are $(1-w)^4$, $(1+w)^4$ and 
$(1-w^2)^2$. But the first two choices are clearly impossible (because 
$\alpha_k(0,0)$ become nonzero constants when $k$ is even or odd), leading to
$\fq(w)=(1-w^2)^2$ which is equivalent to $b=0$, $a=1$. Thus $L=L_0$ 
completing the proof.
\qed

\section*{Acknowledgments} I am grateful to Leonid Dickey and 
George Wilson for helpful discussions and valuable remarks. 
I thank a referee for suggestions that led to an improved  
version of the paper.


\end{document}